\documentclass[prl,preprint,endfloats]{revtex4}

\usepackage{graphicx}
\usepackage{bm}
\usepackage{pifont}
\usepackage{amsmath}

\begin{document}


\title{Propagating spin excitations along skyrmion strings}

\author{S. Seki$^{1,2}$, M. Garst$^3$, J. Waizner$^4$, R. Takagi$^2$, Y. Okamura$^1$, K. Kondou$^2$, F. Kagawa$^{1,2}$, Y. Otani$^{2,5}$, and Y. Tokura$^{1,2}$} 
\affiliation{$^1$ Department of Applied Physics, University of Tokyo, Tokyo 113-8656, Japan, \\ $^2$ RIKEN Center for Emergent Matter Science (CEMS), Wako 351-0198, Japan,  \\ $^3$ Institut f\"ur Theoretische Physik, Technische Universit\"at Dresden, 01062 Dresden, Germany, \\ $^4$ Institut f\"ur Theoretische Physik, Universit\"at zu K\"oln, Z\"ulpicher Str. 77a, 50937 K\"oln, Germany, \\ $^5$ Institute for Solid State Physics, University of Tokyo, Kashiwa 277-8581, Japan}


\begin{abstract}

{\bf Magnetic skyrmions, topological solitons characterized by a two-dimensional swirling spin texture, have recently attracted attention as stable particle-like objects. In a three-dimensional system, a skyrmion can extend in the third dimension forming a robust and flexible string structure, whose unique topology and symmetry are anticipated to host nontrivial functional responses. Here, we experimentally demonstrate the coherent propagation of spin excitations along skyrmion strings for the chiral-lattice magnet Cu$_2$OSeO$_3$. We find that this propagation is directionally non-reciprocal, and the degree of non-reciprocity, as well as the associated group velocity and decay length, are strongly dependent on the character of the excitation modes. Our theoretical calculation establishes the corresponding dispersion relationship, which well reproduces the experimentally observed features. Notably, these spin excitations can propagate over a distance exceeding 10$^3$ times the skyrmion diameter, demonstrating the excellent long-range nature of the excitation propagation on the skyrmion strings. Our combined experimental and theoretical results offer a comprehensive account of the propagation dynamics of skyrmion-string excitations, and suggest the possibility of unidirectional information transfer along such topologically-protected strings.}

\end{abstract}
\maketitle

Recently, the concept of a magnetic skyrmion, i.e. topologically stable spin configuration whose spins point in all of the directions wrapping a sphere, has attracted enormous attention\cite{NeutronMnSi, TEMFeCoSi, SkXReviewFertTwo, SkXReviewTokura}. In a magnetically two-dimensional system, a skyrmion appears as a vortex-like swirling spin texture with particle nature as shown in Fig. 1j. The stable particle nature of the skyrmion suggests its potential application as a novel magnetic bit for future data storage devices, and the nontrivial topology and symmetry of the skyrmion also cause exceptional electromagnetic responses\cite{SkXReviewFertTwo, SkXReviewTokura}.

In three dimensional systems, the skyrmion can form a string structure by extending in the third dimension, which consists of the uniform stacking of two-dimensional skyrmions along the string direction (Fig. 1a)\cite{Monopole, Holography}. Skyrmion strings can be considered as an analogue of the vortex-line in superfluids\cite{RMP_superfluids, PNAS_Kelvinwave}, type-II superconductors\cite{RMP_superconductor} and trapped dilute-gas Bose-Einstein condensates\cite{Science_BEC} or the cosmic string in the universe\cite{cosmic_string}. They are all flexible, and some of these strings are proposed to host a resonant oscillation mode propagating through the string path. This implies the possible coherent signal transfer along skyrmion strings, while the propagation character of their excitations has rarely been investigated before.

Experimentally, such skyrmion strings appear in a series of bulk magnets with chiral cubic atomic lattice. The examples are metallic B20\cite{NeutronMnSi, TEMFeCoSi} or $\beta$-Mn type Co-Zn-Mn\cite{CoZnMn_First} alloys and insulating Cu$_2$OSeO$_3$\cite{Cu2OSeO3_Seki, Cu2OSeO3_SANS_Pfleiderer}, the latter of which is the target of this work. In these materials, the Dzyaloshinskii-Moriya (DM) interaction is the key for the skyrmion formation. For a limited temperature ($T$) range, these compounds host a hexagonal lattice of skyrmion strings aligned along the static magnetic field ($H$) direction (Fig. 1a). This skyrmion crystal (SkX) phase is predicted to host three distinctive magnetic resonance modes\cite{Mochizuki12, B20_FMR, PRB_SkX_spinwave} (central panels in Figs. 1f-h), i.e. the counter-clockwise (CCW) and clockwise (CW) rotational modes both excited by an oscillating magnetic field $H^\nu \perp H$, as well as the breathing (B) mode excited by $H^\nu \parallel H$, which have recently been identified by magnetic resonance experiments\cite{B20_FMR, Onose}. However, these previous works mostly focused on the character of non-propagating uniform excitations with wave number $k^{\rm SW} = 0$. To understand their propagation character, the employment of a different experimental approach sensitive to the $k^{\rm SW} \neq 0$ regime, as well as the theoretical identification of their dispersion relation, is essential.

In this work, we investigated the propagation characteristics of such spin excitations along skyrmion strings. We find that the counter-propagating spin excitations show different propagation behavior, and the degree of non-reciprocity, as well as the associated group velocity and decay length, are strongly dependent on the character of the excitation modes. These experimental features are well reproduced by our theoretically calculated dispersion relations. Moreover, the observed decay lengths exceed 10$^3$ times the skyrmion diameter, reflecting the excellent long-range order of the skyrmion string structure. The present results revealed the peculiar propagation dynamics of skyrmion string excitations, and suggest that skyrmion string can be a good medium for magnon transport with unique functionalities.

The detailed operating principle of the present measurement technique (propagating spin wave spectroscopy) is provided in Refs. \cite{AESWS_doppler_B, AESWS_doppler_Science}. Figures 1b and c indicate the device structure employed in this study, where a plate-shaped Cu$_2$OSeO$_3$ single crystal with thickness $b \approx 2\mu$m is placed on top of a pair of Au coplanar waveguides (CPW) fabricated on a Si substrate. When the oscillating electric current of frequency $\nu$ is injected into a CPW, the generated oscillating magnetic field $H^\nu$ induces resonant spin excitations in the neighboring Cu$_2$OSeO$_3$ sample as shown in Fig. 1c. These spin excitations propagate along the sample and induce an oscillating electric voltage in the CPWs through the inverse process. By measuring the magnetic contribution to the self-inductance $\Delta L_{11}$ and mutual inductance $\Delta L_{nm}$ spectra (with $m$ and $n$ representing the port number used for the excitation and detection, respectively) for CPWs using a vector network analyzer (NA), we can directly evaluate the local excitation character and propagation character of spin excitations, respectively. Unless specified, we employed a device structure with $\lambda^{\rm SW} = 12 \mu$m and $d=20 \mu$m for all the measurements; here, $\lambda^{\rm SW}$ and $d$ are the period of a CPW pattern and the gap distance between two CPWs as shown in Figs. 1c and S3a, respectively. The wave number distribution of the induced spin excitations is determined by the Fourier transform of CPW pattern (Fig. S2)\cite{AESWS_doppler_B}, whose maximum intensity is always located at $k^{\rm SW} = 2\pi/\lambda^{\rm SW}\approx 0.5 \mu m^{-1}$. The propagation direction of $k^{\rm SW}$ and $H$ are set parallel to the [110] direction of the Cu$_2$OSeO$_3$ crystal.

Figure 1d indicates the $H$-$T$ magnetic phase diagram for Cu$_2$OSeO$_3$, obtained for the field cooling path shown by the arrows. While the SkX phase usually appears as the thermodynamic equilibrium state only for the narrow temperature range just below $T_c$ (Fig. S1)\cite{Cu2OSeO3_Seki, Cu2OSeO3_SANS_Pfleiderer}, the rapid cooling ($\approx 5$K/min) of the sample at 17.5 mT enabled us to keep the SkX phase as a metastable state\cite{Quench} even down to 10 K. The corresponding $H$-dependence of local magnetic resonance spectra $\Delta L_{11}$ measured at 25 K is plotted in Fig. 1e, where the pure SkX phase characterized by the three magnetic resonance modes\cite{Mochizuki12, B20_FMR} is clearly identified between 20 mT and 47.5 mT.

Next, we investigated the propagation characteristics of these spin excitations in the SkX phase. Figures 2a-c indicate the spectra of mutual inductance $\Delta L_{21}$ and $\Delta L_{12}$ measured at 25 K and +25 mT, which represent the propagation character of spin excitations with the wavevector $+k^{\rm SW}$ and $-k^{\rm SW}$ (i.e., parallel and anti-parallel to the external magnetic field as shown in Fig. 2d), respectively. Here, Figs. 2a-c correspond to the CCW, breathing and CW modes as schematically illustrated in Figs. 1f-h, respectively. For all modes, a propagating signal of coherent spin excitations is observed, implying that the skyrmion strings are well-defined and free of defects over an extended distance. Importantly, for the CCW mode, we can identify a clear frequency shift $\Delta \nu$ between $\Delta L_{21}$ and $\Delta L_{12}$, which demonstrates that the spin excitations propagating along the positive and negative direction on the skyrmion strings are not equivalent. On the other hand, the magnitude of $\Delta \nu$ is relatively small for the breathing and CW modes. When the direction of external $H$ is reversed, the sign of $\Delta \nu$ is reversed as shown in Figs. 2e-h. Such a nonreciprocal behavior and its mode dependence can be interpreted in terms of the dispersion relation as discussed below. 

Figure 2i shows the spin excitation dispersion in the SkX phase, theoretically calculated for the present $k^{\rm SW}\parallel H$ configuration (see supplementary information for details). The dispersion relations are asymmetric (i.e. the resonance frequency is not an even function of $k^{\rm SW}$) for all three modes, and this asymmetry is most pronounced for the CCW mode. Here, the experimentally measured $\Delta \nu$ corresponds to the difference of eigen frequencies between $k^{\rm SW}=\pm 2\pi/\lambda^{\rm SW}$, which directly reflects the degree of asymmetry in the dispersion relation. Therefore, the observation of the largest magnitude of $\Delta \nu$ in the CCW mode is in accord with the predicted dispersions in Fig. 2i.

In the following, we discuss the microscopic origin of the observed nonreciprocity and its mode dependence. For skyrmion strings, we find that there are two contributions to $\Delta \nu$: the first is attributed to the DM interaction, $\Delta \nu_{\rm DM}$, and the second arises from the dynamic dipolar interaction, $\Delta \nu_{\rm dip}$. We refer for a discussion of the latter to the supplementary information, and focus here on the contribution from the DM interaction\cite{Kataoka, SW_DM_BLS, LiFe5O8, SW_DM_COSO, SW_DM_MnSi}. Previously, it has not systematically been understood how DM interaction affects the dispersion relation in the noncollinear magnets with arbitrary excitation modes. As detailed below, we derive the general analytic expression of $\Delta \nu$, which reveals that the spatial distribution of local precession amplitude, as well as the static spin texture in the ground state, plays an important role to define the degree of dispersion asymmetry in each mode.

Figures 1f-h represent snapshot images describing how the local spin excitations launched at the $z=0$ plane propagates on a skyrmion string (aligned along the $H \parallel z$ direction) with the wave vector $+k^{\rm SW}$ and $-k^{\rm SW}$. In the right panel, the corresponding cross-sectional images for selected $z$-planes are also indicated. Such a propagating spin excitation induces a change of the DM energy density $\varepsilon_{\rm DM} = {\bf D}_z ({\bf m}_i \times {\bf m}_j)$, which is defined between the local moments ${\bf m}_i$ and ${\bf m}_j$ on adjacent sites $i$ and $j$ along the ${\hat z}$-direction with the DM vector ${\bf D}_z = D \hat z$. At the edge region of the skyrmion, the local moment is pointing parallel to ${\bf H}$ in the ground state, and precesses counter-clockwise around ${\bf H}$ in the excited state as shown in Fig. 1i. In this case, the $+k^{\rm SW}$ and $-k^{\rm SW}$ excitation modes locally induce a conical spin arrangement along the ${\hat z}$-direction but with opposite sign of the spin helicity $({\bf m}_i \times {\bf m}_j)$ and thus of $\varepsilon_{\rm DM}$, which leads to a different excitation energy for modes with $\pm k^{\rm SW}$ and causes an asymmetric dispersion. At the core region of the skyrmion, on the other hand, the local moment is antiparallel to ${\bf H}$ and precesses clockwise around ${\bf H}$; therefore, the sign of $\varepsilon_{\rm DM}$ contribution is reversed as compared with the skyrmion edge position (Fig. 1i). As a consequence, the magnitude of non-reciprocity deriving from the DM interaction is basically determined by the imbalance of local precession amplitudes between the edge and core positions with local moments pointing along $+\hat z$ and $-\hat z$ directions, respectively.

More precisely, $\Delta \nu_{\rm DM}$ due to the DM interaction in the limit of small $k^{\rm SW}$ is given by
\begin{equation} \label{Nreci_DM}
\Delta \nu_{\rm DM} \propto D |k^{\rm SW}| \int_{\rm u.c.} d^2{\bf r}\, \hat z\,i 
(\delta {\bf m}({\bf r}) \times \delta {\bf m}^*({\bf r}))
\propto D |k^{\rm SW}| \int_{\rm u.c.} d^2{\bf r}\, m^z_0({\bf r}) \mathcal{A}({\bf r}),
\end{equation}
with the integration range defined by the two-dimensional magnetic unit cell of the SkX. Here, we assume the local magnetization dynamics ${\bf m}({\bf r})={\bf m}_0 ({\bf r}) + (\delta {\bf m}({\bf r}) \exp[i 2\pi \nu t] + {\rm c.c.})$, with ${\bf m}_0 ({\bf r})$ and ${\bf \delta m} ({\bf r})$ representing the static and dynamical parts of local magnetization component, respectively; $m^z_0 ({\bf r})$ is ${\hat z}$-component of ${\bf m}_0 ({\bf r})$. The cross product in the integrand is proportional to the local precession intensity $\mathcal{A}({\bf r})$ as defined in Fig. 1n (i.e. the area enclosed by the precessing local magnetization during a single oscillation period, that quantifies the probability density of the magnon). This allows for an intuitive geometric interpretation of $\Delta \nu_{\rm DM}$ as given by the last term in Eq.~\eqref{Nreci_DM}. Its integrand is determined by the product of $m^z_0({\bf r})$ and precession density $\mathcal{A}({\bf r})$ for each excitation mode, whose spatial distribution are presented in Fig. 1j and Figs.~1k-m, respectively. In case of the CCW mode (Fig. 1k), the precession density is confined to the edge of the unit cell, resulting in a large magnitude of $\Delta \nu_{\rm DM}$ according to Eq.~\eqref{Nreci_DM}. For the CW mode (Fig. 1m), in contrast, a finite precession density is present both at the core and the edge regions, which leads to a cancelation for $\Delta \nu_{\rm DM}$. The breathing mode (Fig.~1l) is characterized by a small precession density at the core and the edge region and therefore hosts only a tiny $\Delta \nu_{\rm DM}$. This analysis accounts already qualitatively for the large $\Delta \nu$ observed for the CCW mode, and demonstrates that the spatial details of the internal magnetic texture as well as its excitation manner plays a crucial role for the magnitude of non-reciprocity. As we explain in the supplementary information, for a quantitative comparison an additional contribution $\Delta \nu_{\rm dip}$ attributed to the dynamical stray field must be taken into account.

The preceding discussion focused on the dispersion relation in the bulk limit $k^{\rm SW} \gg 1/b$. For $k^{\rm SW} \lesssim 1/b$, the magnetostatic (i.e. magnetic dipolar) interaction results in an additional peak structure in the dispersion whose sharpness scales with the sample thickness $b$\cite{MSW_Text}. Our present experimental setup corresponds to $k^{\rm SW} \approx 1/b$. As sketched in Fig. 2j and k, this magnetostatic dispersion (solid lines) develops on the background of the non-reciprocal bulk spectrum (dashed lines) so that $\Delta \nu$ is still dominated by the latter. Nevertheless, the group velocity $v_g = 2\pi (\partial \nu/\partial k^{\rm SW})$ is strongly influenced by the magnetostatic interactions for $k^{\rm SW} \lesssim 1/b$. Experimentally, the magnitude of group velocity can be deduced based on the relationship $|v_g| = \nu_{pp} d$\cite{AESWS_doppler_B}, with $\nu_{pp}$ representing the oscillation period of $\Delta L_{nm}$ signals as shown in Fig. 2b.

Figures 3a-c indicate the magnetic field dependence of magnetic resonance frequency $\nu_0$, frequency shift $\Delta \nu$ between $\pm k^{\rm SW}$, and magnitude of group velocity $|v_g|$, experimentally deduced at 25 K from the similar data sets as shown in Figs. 2a-c. Theory provides parameter-free predictions for these quantities that are also plotted in Figs. 3d-f. In both cases, the non-reciprocal frequency shift $\Delta \nu$ is much larger for the CCW mode than for the other modes, and the magnitude of $|v_g|$ is the largest and smallest for the breathing and CW modes, respectively. The order of $\Delta \nu$ and $|v_g|$ are also roughly in accord with each other. Such a good agreement between the experimental and theoretical results firmly establishes the overall picture of dispersion relations for skyrmion string excitations.

On the basis of the present experimental data, we can further deduce the decay length $l$ of propagating spin excitations. Figure 4a indicates the amplitude spectra of self inductance $|\Delta L_{11}|$ and mutual inductance $|\Delta L_{21}|$ measured at 25 K and +25 mT, whose ratio provides the decay rate of spin excitation amplitude during the propagation over the distance $d$. The decay length $l$ can be estimated from the relationship $2 |\Delta L_{21}|/|\Delta L_{11}| = \exp (-d/l)$, which is associated with the damping parameter $\alpha$ in the form of $l = v_g / (2\pi \nu_0 \alpha)$\cite{AESWS_doppler_B}. In Figs. 4b and c, the magnetic field dependence of $l$ and $\alpha$ at 25 K are plotted. While the decay length $l$ largely depends on the character of spin excitation modes, basically the same value of $\alpha$ is shared by all three modes in the SkX phase. Here, the CW mode hosts considerably shorter $l$ than the other two modes, reflecting its large $\nu_0$ and small $v_g$ values (Figs. 3a and c). In Figs. 4d and e, temperature dependence of $l$ and $\alpha$ for various spin excitation modes are plotted. For all temperatures, the SkX phase is characterized by a damping parameter $\alpha$ that is slightly larger, but less than a factor of two, than the one in the field-induced ferromagnetic phase. This slight enhancement might arise from the elliptical nature of local magnetization precession\cite{dampingEnhance} and/or the additional scattering, e.g., from defects of the SkX order. Recently, it was suggested that the SkX contains a considerable amount of singular point defects interpreted as emergent magnetic monopoles\cite{Monopole}, which act like the slider of a zipper connecting two skyrmion strings. The existence of such monopole-like defects diminishes the spatial long-range order of the SkX and thus also the coherent propagation of spin excitation along the skyrmion strings. Nevertheless, the decay length $l$ in the SkX phase still exceeds 50 $\mu$m at low temperatures, proving that the spin excitations on the skyrmion strings can propagate a distance exceeding 10$^3$ times the diameter of single skyrmion string ($\approx 50$ nm\cite{Cu2OSeO3_Seki, Cu2OSeO3_SANS_Pfleiderer}). Such a small damping and long propagation distance of the excitation are rather unexpected for the SkX phase with its intricate three-dimensional spin texture, and the above results suggest that the skyrmion string with excellent long-range order can be a good medium for magnon transport.

 In this study, we experimentally demonstrated the coherent and nonreciprocal propagation of spin excitations along skyrmion strings. Unlike conventional ferromagnets, skyrmion strings host multiple excitation modes, and their group velocity, decay length, and degree of nonreciprocity turned out to be strongly mode-dependent in full agreement with our calculated dispersion relations. These results establish the comprehensive picture of the propagation dynamics of skyrmion string excitations. Moreover, we developed a general theoretical framework to evaluate the dispersion asymmetry for non-collinear magnets, which reveals that the underlying static spin texture and spatial distribution of local precession amplitude are key factors that determine the degree of nonreciprocity for each mode. Since the propagating spin excitations carry energy and momentum, the above findings suggest the possibility of unidirectional information transfer along skyrmion strings.
 
 On a broader perspective, the interesting extension of the present work includes the study of the skyrmion string dynamics in more general conditions. While the current experiments focused on straight skyrmion strings embedded in the skyrmion crystals, the dynamics of a single isolated string has also been discussed theoretically very recently\cite{isolatedStringOne, isolatedStringTwo, isolatedStringThree}. Skyrmion strings can be bent into curved shapes due to their topological protection\cite{Monopole}, and the associated spin excitations are expected to propagate along the string path in analogy with the Kelvin mode of vortex lines in superfluids\cite{RMP_superfluids, PNAS_Kelvinwave, isolatedStringTwo}. These features suggest that skyrmion strings may serve as robust and flexible information transmission lines without Joule heat loss. The nonlinear dynamics and networking of these strings are another fascinating issues. Our results highlight that not only the skyrmion particle in two-dimensional systems, but also the skyrmion string in three-dimensional systems is an attractive topological object, and further investigation of its fundamental and functional properties will be a promising endeavor.

\begin{figure}
\begin{center}
\includegraphics*[width=17cm]{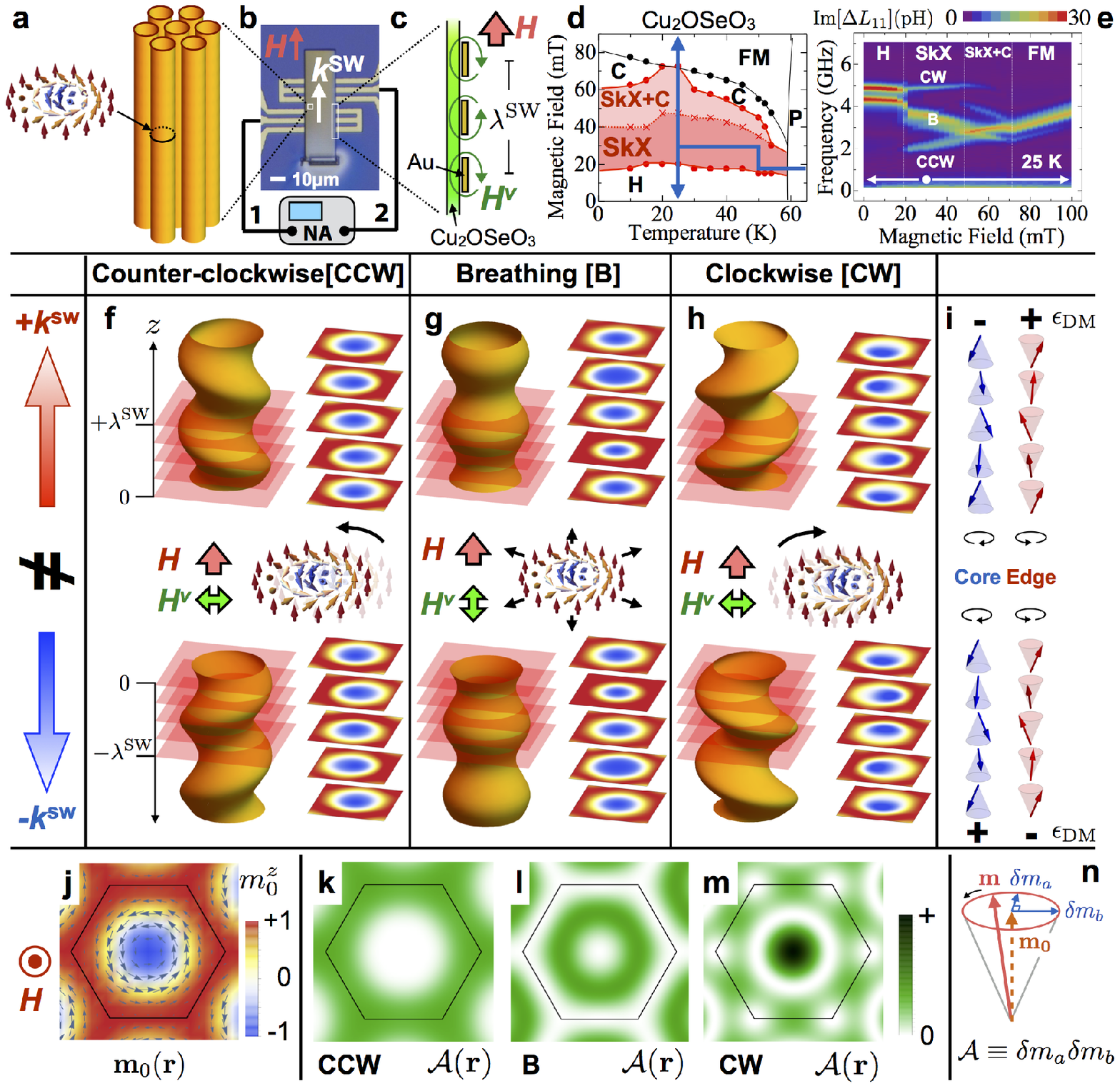}
\caption{{\bf Propagating excitation modes on skyrmion strings.} {\bf a}, Schematic illustration of skyrmion strings. {\bf b,c}, Top view optical image of device structure and side view illustration of a coplanar waveguide, where the AC current injected from network analyzer (NA) generates oscillating magnetic field $H^\nu$ and causes spin excitation in the neighboring Cu$_2$OSeO$_3$. {\bf d}, $H$-$T$ magnetic phase diagram for Cu$_2$OSeO$_3$, obtained with the cooling path shown by the arrows. SkX, C, H, FM, and P represent the skyrmion lattice, conical, helical, ferromagnetic, and paramagnetic states, respectively. {\bf e}, The corresponding $H$-dependence of magnetic resonance spectra $\Delta L_{11}$ at 25 K. {\bf f}-{\bf h}, Schematic illustration of CCW, breathing and CW excitation modes on skyrmion strings. The central part represents the local oscillation manner of skyrmion at the $z=0$ plane. The upper and lower parts are the snapshot images describing how the spin excitation launched at $z=0$ propagates on the skyrmion strings, along the $\pm z$ direction parallel and antiparallel to $H$, respectively. The cross sectional images describing the size and position of skyrmion at selected $z$-planes (shown by red layers) are also indicated. {\bf i}, The direction of local magnetic moment at the core and edge position of skyrmion in each $z$ layer. Black rounded arrows denote the sense of local moment precession in the time domain, and +/- symbols indicate the sign of local DM energy gain $\epsilon_{\rm DM}$. {\bf j}-{\bf m}, Calculated spatial distribution of local magnetization direction ${\bf m}_0 ({\bf r})$ in the ground state, and local amplitude of elliptical magnetization precession $\mathcal{A} ({\bf r})$ for the CCW, B and CW excitation modes in the SkX phase. In {\bf j}, the arrows and background color ($m_0^z ({\bf r})$) represent the in-plane and out-of-plane component of ${\bf m}_0 ({\bf r})$, respectively. {\bf n}, Schematic illustration of local magnetization dynamics, described by ${\bf m}={\bf m_0}+\sqrt{2}{\rm Re}[(\delta m_a {\bf e_a} + i\delta m_b {\bf e_b}) \exp[i2\pi \nu t]]$ with ${\bf e_a}$ and ${\bf e_b}$ being the orthogonal unit vectors normal to ${\bf m_0}$. Here, we define the local elliptical precession amplitude $\mathcal{A} \equiv \delta m_a \delta m_b$, whose spatial distribution for each mode is plotted in {\bf k}-{\bf m}.}
\end{center}
\end{figure}

\begin{figure}
\begin{center}
\includegraphics*[width=16cm]{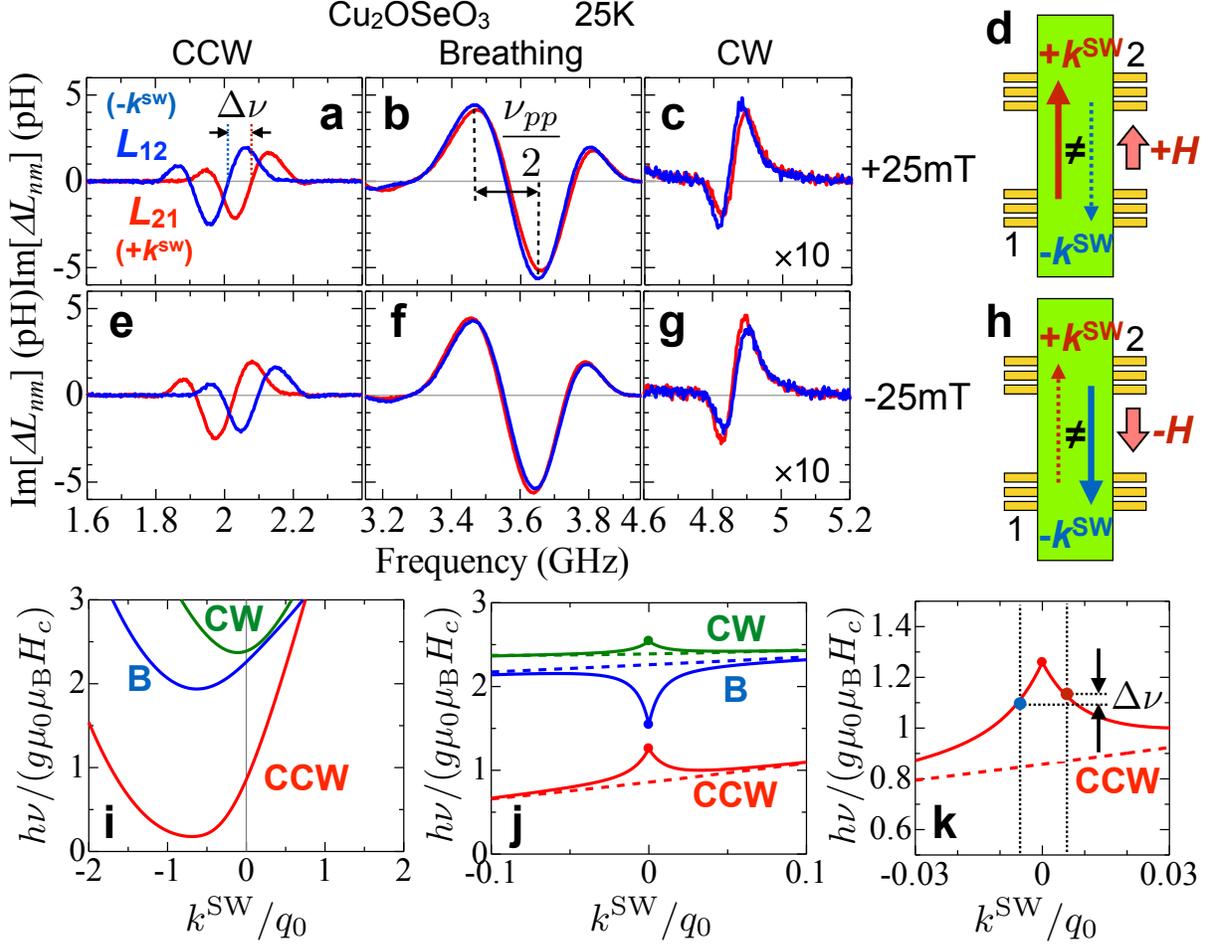}
\caption{{\bf Propagation character of spin excitations on skyrmion strings.} {\bf a}-{\bf c}, The spectra of mutual inductance $\Delta L_{21}$ and $\Delta L_{12}$, which represent the propagation character of spin excitation with the wave vector $+k^{\rm SW}$ and $-k^{\rm SW}$, respectively. The data were measured in the SkX state under the configuration shown in {\bf d} at 25 K with $\mu_0 H = +25$ mT. {\bf a}, {\bf b}, and {\bf c} represent the CCW, B and CW modes, respectively. {\bf e}-{\bf h}, The corresponding data measured with $\mu_0 H = -25$ mT. {\bf i}-{\bf k}, Dispersion relation for various spin excitations on skyrmion strings, theoretically calculated for the $k^{\rm SW} \parallel H$ configuration with $H=0.4 H_c$. Here, $h$, $g$, $\mu_0$, $\mu_{\rm B}$, $H_c$ and $q_0$ are Planck constant, $g$-factor, vacuum magnetic permeability, Bohr magneton, critical field value to induce FM state and magnetic modulation pitch, respectively. In {\bf j} and {\bf k}, the solid and dashed lines represent the dispersion relation for a finite sample in the magnetostatic limit and an infinitely extended sample in the bulk limit, respectively.}
\end{center}
\end{figure}

\begin{figure*}
\begin{center}
\includegraphics*[width=15cm]{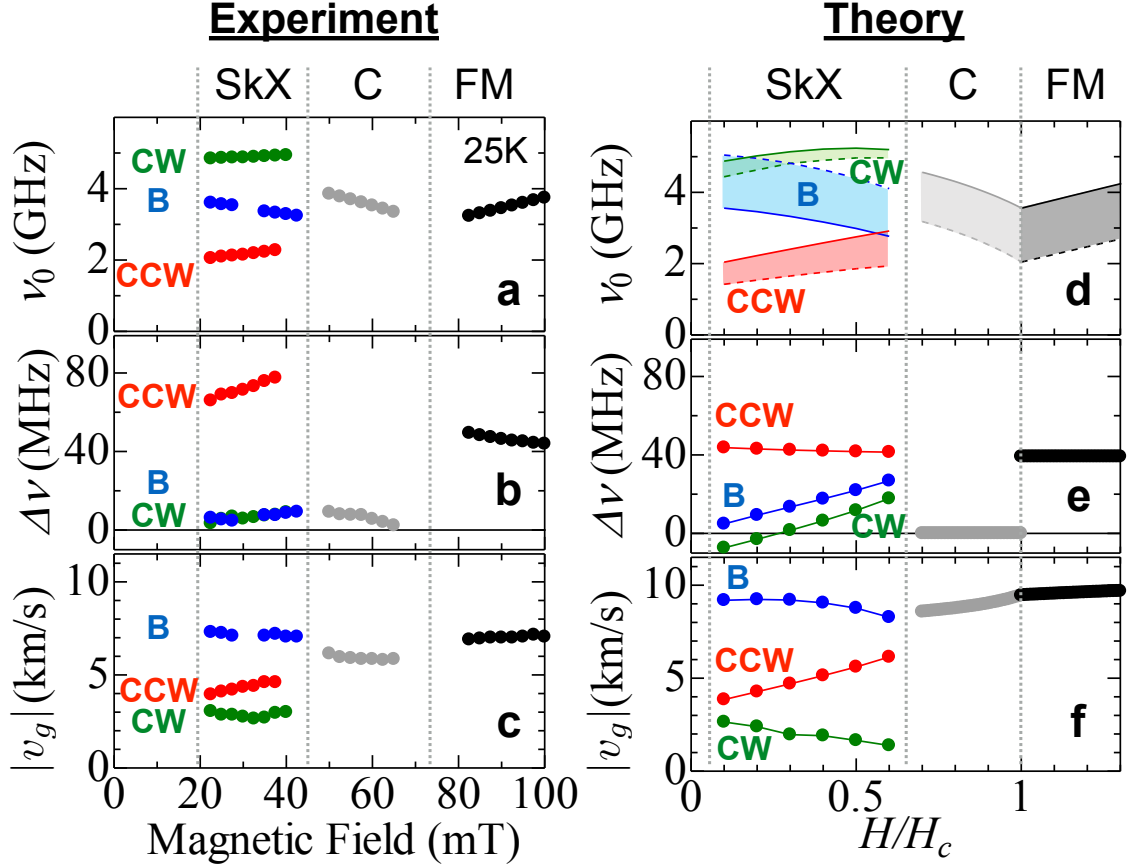}
\caption{{\bf Magnetic field dependence of propagation character for various spin excitation modes.} {\bf a}-{\bf c}, Experimentally obtained magnetic field dependence of ({\bf a}) magnetic resonance frequency $\nu_0$, ({\bf b}) frequency shift $\Delta \nu$ between $\pm k^{\rm SW}$, and ({\bf c}) magnitude of group velocity $|v_g|$ at 25 K. Note that the data for the conical spin state is taken in the $H$-increasing run after the zero field cooling (Fig. S1). {\bf d}-{\bf f}, The corresponding data calculated from the theoretical dispersion relation. In {\bf d}, the $k^{\rm SW} \rightarrow 0$ values in the bulk and magnetostatic limit are plotted as dashed and solid lines, respectively, whose difference $\delta \nu_{\rm d}$ provides an estimate of the group velocity $v_g \approx 2\pi b (\delta \nu_{\rm d})$ shown in {\bf f}. In {\bf e}, the non-reciprocal frequency shift $\Delta \nu$ is shown computed for $|k^{\rm SW}| = 0.5 \mu m^{-1}$.}
\end{center}
\end{figure*}

\begin{figure}
\begin{center}
\includegraphics*[width=15cm]{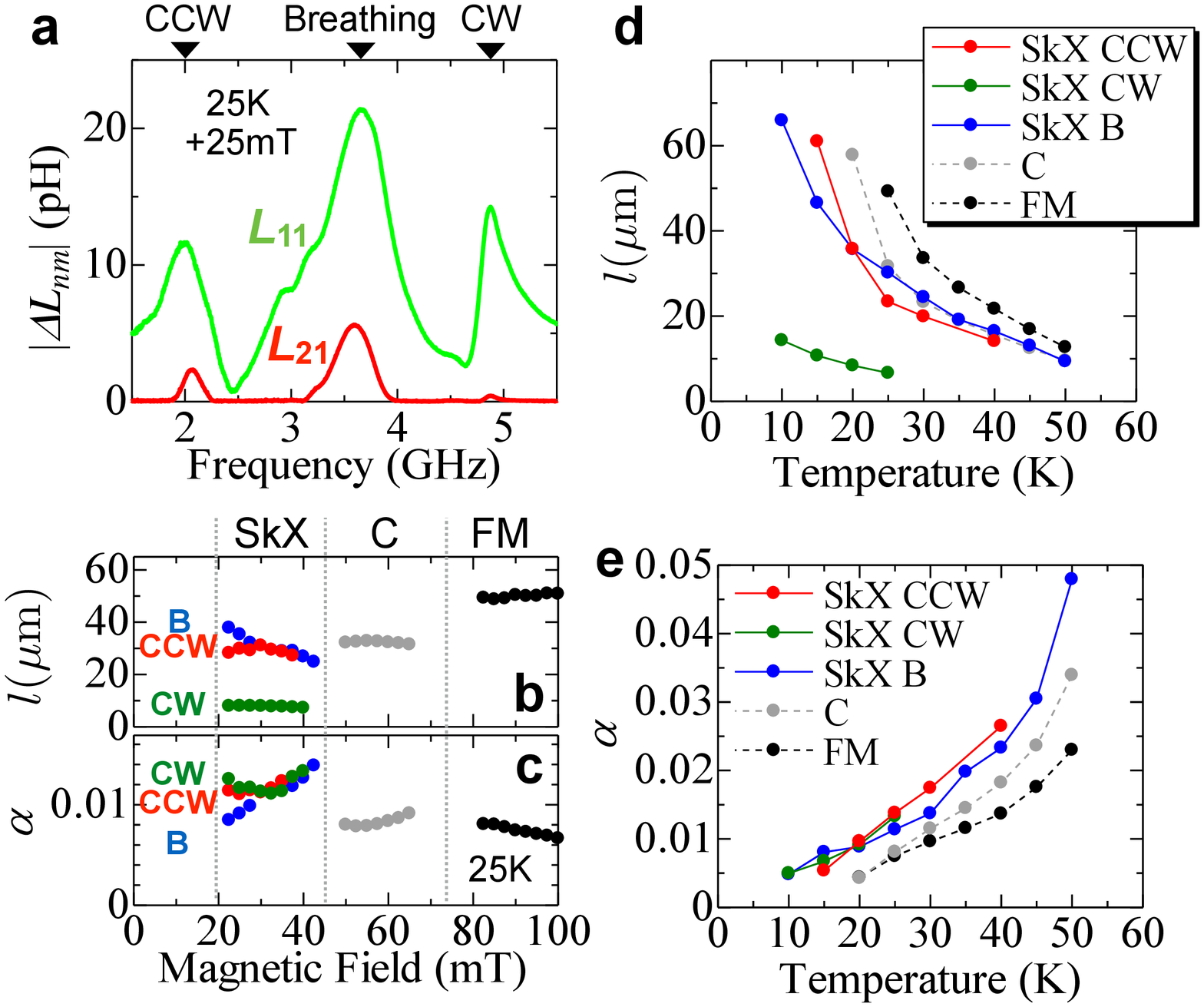}
\caption{{\bf Decay length and damping parameter associated with the propagation of spin excitations.} {\bf a}, Amplitude spectra of self inductance $|\Delta L_{11}|$ and mutual inductance $|\Delta L_{21}|$ measured in the SkX state at 25 K and +25mT, whose ratio gives the decay rate of spin excitation amplitude during the propagation. {\bf b,c}, Magnetic field dependence of decay length $l$ and damping parameter $\alpha$ measured at 25 K. {\bf d,e}, Temperature dependence of decay length and damping parameter for various spin excitation modes.} 
\end{center}
\end{figure}

\section*{Methods} 

Single crystals of Cu$_2$OSeO$_3$ were grown by the chemical vapor transport method. A pair of Au coplanar wave guides (CPW: 200 nm thickness) were fabricated on the oxidized silicon substrate through the standard photolithography technique, and a plate-shaped single crystal of Cu$_2$OSeO$_3$ ($\sim 2 \mu$m thickness) was placed across them with W (tungsten) deposition at an edge of the crystal using the focused ion beam (FIB) micro fabrication technique (Fig. 1b). By measuring the magnetic contribution to the complex spectra of self-inductance $\Delta L_{11}$ and mutual inductance $\Delta L_{nm}$ (with $m$ and $n$ being the port number used for the excitation and detection, respectively) for these CPWs with the vector network analyzer, the local excitation character and propagation character of spin excitation can be directly evaluated, respectively\cite{AESWS_doppler_B, AESWS_doppler_Science}. The spin excitation contribution to the inductance spectrum $\Delta L_{nm} (\nu) = L_{nm}(\nu)-L_{nm}^{\rm ref} (\nu)$ is derived by the subtraction of the common background $L_{nm}^{\rm ref} (\nu)$ from the raw data $L_{nm} (\nu)$. Here, $L_{nm} (\nu)$ taken at $\mu_0 H = 250$ mT is adopted as the reference spectrum $L_{nm}^{\rm ref} (\nu)$, where the magnetic resonance is absent within our target frequency range of 0.2 GHz $\leq \nu \leq$ 7.0 GHz. Unless specified, we employed the device with the central wave length $\lambda^{\rm SW} = 12 \mu$m and the gap distance $d=20 \mu$m.

\section*{Author contributions} S.S. performed the measurements. S.S., R.T., Y.O., K.K., F.K., Y.O. contributed to the crystal growth and device fabrication. M.G. and J.W. performed theoretical calculation. S.S. M.G. and Y.T. planned the project and wrote the manuscript. All authors discussed the results and commented on the manuscript.

\section*{Acknowledge} The authors thank J. Iwasaki, S. Hoshino, N. Nagaosa, and H. Hosono for enlightening discussions and experimental helps. This work was partly supported by Grants-In-Aid for Scientific Research (A) (Grant No. 18H03685) and Grant-in-Aid for Scientific Research on Innovative Area, "Nano Spin Conversion Science" (Grant No.17H05186) from JSPS, and PRESTO (Grant No. JPMJPR18L5) from JST. M. G. was supported by the DFG via SFB 1143 "Correlated magnetism: from frustration to topology" and grant GA 1072/6-1.



\end{document}


\title{Supplementary information:\\Propagating spin excitations along skyrmion strings}

\author{S. Seki$^{1,2}$, M. Garst$^3$, J. Waizner$^4$, R. Takagi$^2$, Y. Okamura$^1$, K. Kondou$^2$, F. Kagawa$^{1,2}$, Y. Otani$^{2,5}$, and Y. Tokura$^{1,2}$} 
\affiliation{$^1$ Department of Applied Physics, University of Tokyo, Tokyo 113-8656, Japan, \\ $^2$ RIKEN Center for Emergent Matter Science (CEMS), Wako 351-0198, Japan,  \\ $^3$ Institut f\"ur Theoretische Physik, Technische Universit\"at Dresden, 01062 Dresden, Germany, \\ $^4$ Institut f\"ur Theoretische Physik, Universit\"at zu K\"oln, Z\"ulpicher Str. 77a, 50937 K\"oln, Germany, \\ $^5$ Institute for Solid State Physics, University of Tokyo, Kashiwa 277-8581, Japan}


\maketitle

\section{Experimental details}

\subsection{Thermodynamic equilibrium phase diagram of Cu$_2$OSeO$_3$}

\begin{figure}[b]
\begin{center}
\includegraphics*[width=14cm]{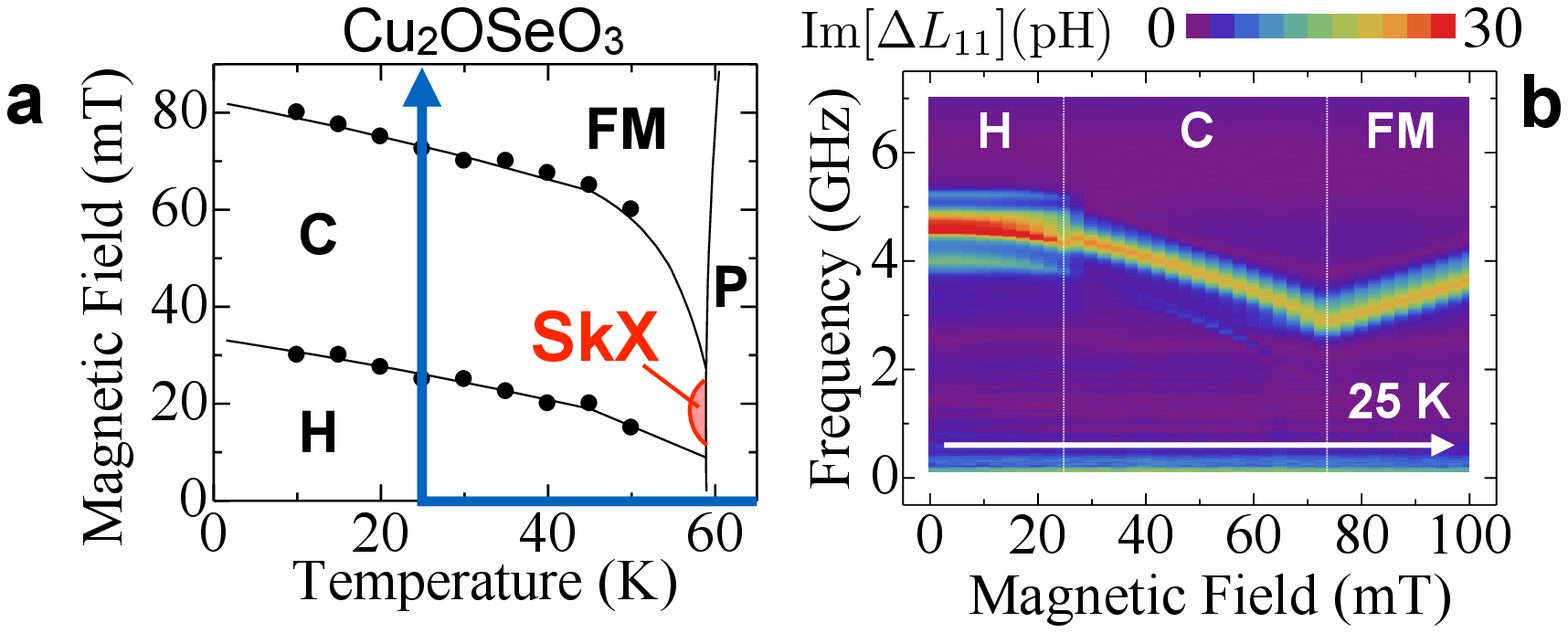}
\caption{{\bf a}, $H$-$T$ magnetic phase diagram for Cu$_2$OSeO$_3$, obtained with the zero field cooling procedure as shown by the arrow. SkX, C, H, FM, and P represent the skyrmion lattice, conical, helical, ferromagnetic, and paramagnetic states, respectively. {\bf b}, The corresponding $H$-dependence of magnetic resonance spectra $\Delta L_{11}$ at 25 K. The data is taken for the $H$-increasing process after zero field cooling.}
\label{ZFC_PhaseDiagram}
\end{center}
\end{figure}

Figure \ref{ZFC_PhaseDiagram}a indicates $H$-$T$ magnetic phase diagram for Cu$_2$OSeO$_3$, obtained with the zero field cooling procedure followed by the $H$-increasing scans. This represents the thermodynamic equilibrium phase diagram, where the SkX phase appears only for the narrow $T$-$H$ region just below $T_c$\cite{Cu2OSeO3_Seki, Cu2OSeO3_SANS_Pfleiderer}. The corresponding $H$-dependence of magnetic resonance spectra $\Delta L_{11}$ measured at 25 K are plotted in Fig. \ref{ZFC_PhaseDiagram}b. From the comparison with the theoretically predicted behaviors\cite{B20_FMR}, we can identify the pure conical spin state between 30 mT and 70 mT. In Figs. 3a-c in the main text, the data sets for the conical spin state were measured through such a zero field cooling procedure.

\subsection{Dependence on the wave number of spin excitation}

\begin{figure}[b]
\begin{center}
\includegraphics*[width=14cm]{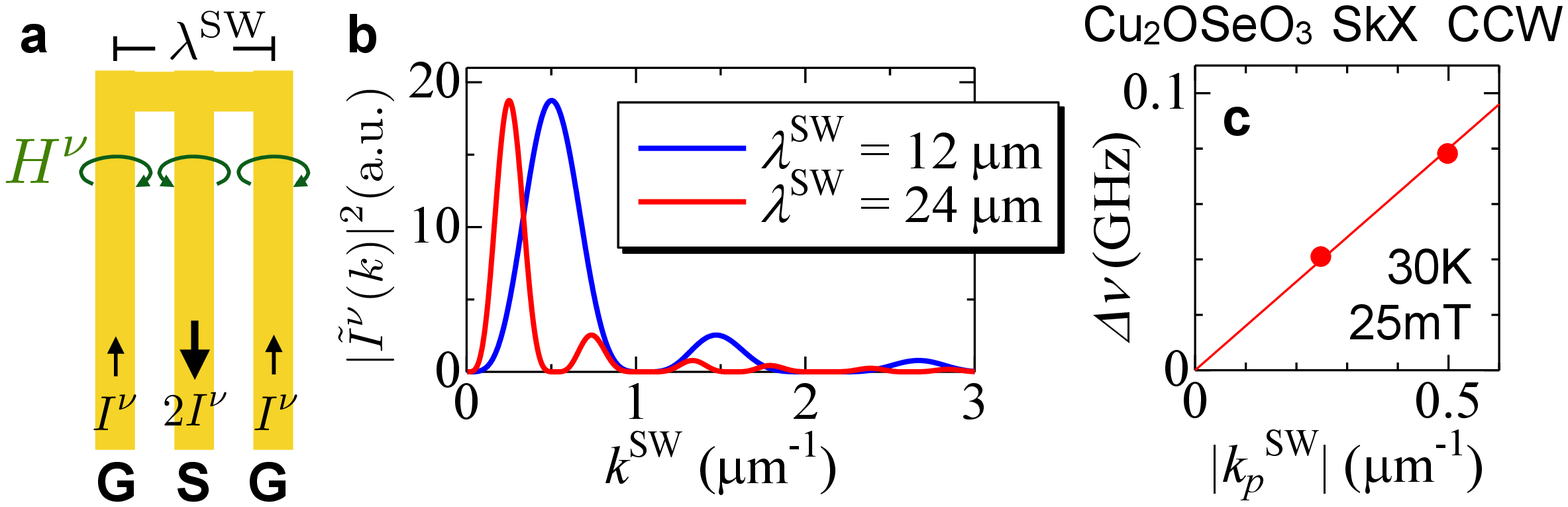}
\caption{{\bf a}, Schematic illustration of coplanar waveguide pattern used for the present study, which consists of a signal (S) line and a pair of ground (G) lines. The oscillating electric current $I^\nu$ injected from network analyzer generates oscillating magnetic field $H^\nu$. Here, G and S lines and the space between them have the width of $\lambda^{\rm SW}/4$. {\bf b}, Fourier transform of the current distribution $|I^\nu (k)|$ for the waveguide pattern shown in {\bf a}, calculated with various $\lambda^{\rm SW}$ values. Here, the maximum peak intensity always appears at $k_p^{\rm SW} = 2\pi/\lambda^{\rm SW}$. {\bf c}, $|k_p^{\rm SW}|$-dependence of frequency shift $\Delta \nu$ between $\pm k^{\rm SW}$, measured for the CCW mode in the SkX state at 30K and +25mT using two different waveguide patterns considered in {\bf b}. The definition of $\Delta \nu$ is given in Fig. 2a in the main text.}
\label{WavenumberDep}
\end{center}
\end{figure}

In the present study, the CPW pattern as shown in Fig. \ref{WavenumberDep}a is employed for the generation and detection of spin excitation. The injection of oscillating electric current $I^\nu$ into a CPW generates oscillating magnetic field $H^\nu$ via the Biot-Savart law, which couples to the spin excitation in the neighboring Cu$_2$OSeO$_3$ crystal. Here, the wave number distribution of induced spin excitation is given by the Fourier transform of the spatial distribution of electric current density\cite{AESWS_doppler_B, AESWS_doppler_Science}. In Fig. \ref{WavenumberDep}b, the corresponding profiles for the CPW pattern in Fig. \ref{WavenumberDep}a with the period of $\lambda^{\rm SW} = 12 \mu$m and $\lambda^{\rm SW} = 24 \mu$m are plotted. In both cases, the maximum peak intensity appears at $k_p^{\rm SW} = 2\pi/\lambda^{\rm SW}$. Since the intensity of higher order peak is almost one order of magnitude smaller, we analyzed our $\Delta L_{nm}$ data assuming that the contribution from the main peak centered at $k_p^{\rm SW}$ is dominant. 

Note that Eq. (1) in the main text predicts the relationship $\Delta \nu \propto |k_p^{\rm SW}|$, which suggests that the magnitude of frequency shift $\Delta \nu$ between $\pm k^{\rm SW}$ (as observed in Fig. 2a in the main text) should depend on the $|k_p^{\rm SW}|$ value of the CPW pattern. In Fig. \ref{WavenumberDep}c, the  $\Delta \nu$ value experimentally measured for the CCW mode in the SkX state is plotted as a function of $|k_p^{\rm SW}|$. This data confirms the predicted $\Delta \nu \propto |k_p^{\rm SW}|$ relationship, in accord with the asymmetric dispersion as shown in Figs. 2i-k in the main text.

\subsection{Dependence on the gap distance}

\begin{figure*}[b]
\begin{center}
\includegraphics*[width=14cm]{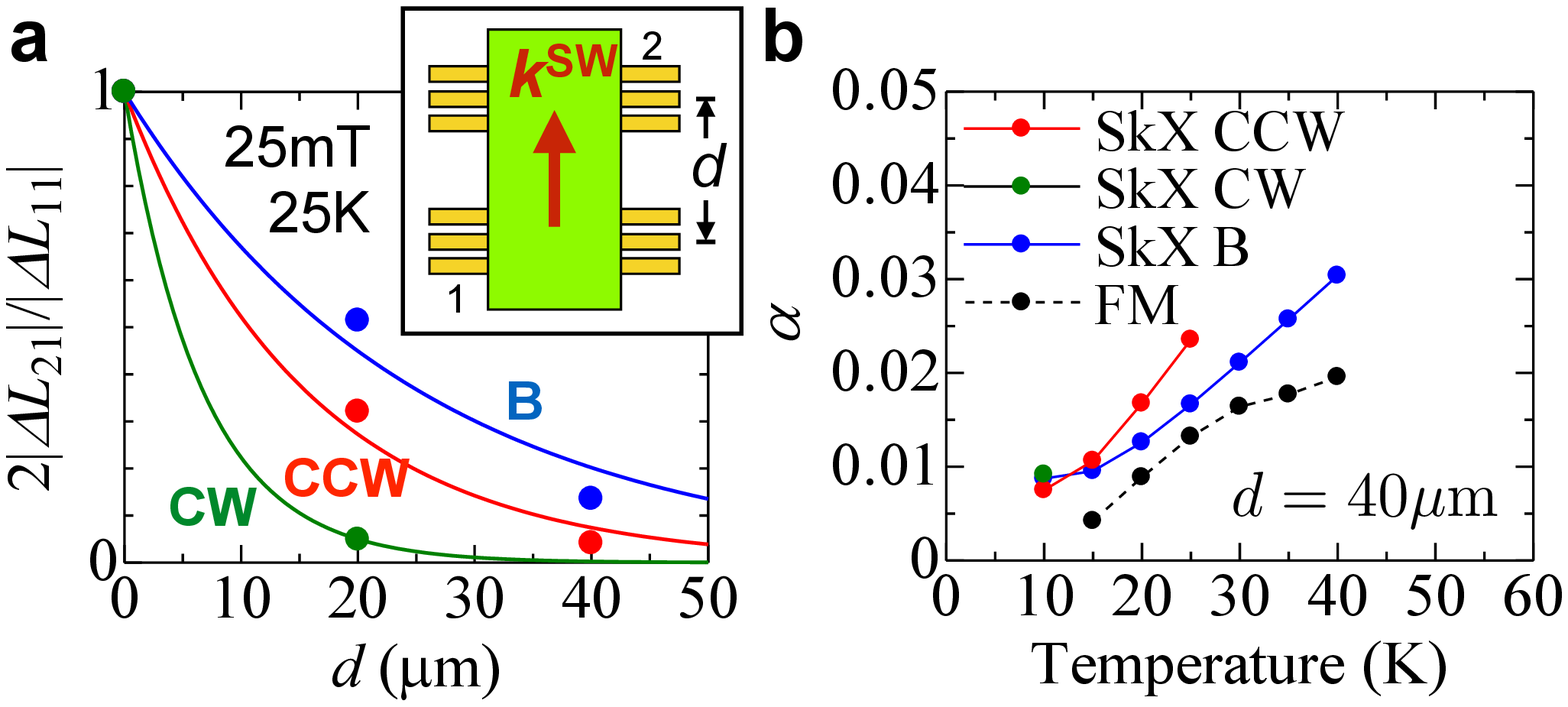}
\caption{{\bf a}, The decay rate of spin excitation amplitude during the propagation, plotted as a function of the gap distance $d$. The measurement is performed for various spin excitation modes in the SkX state at 25K and 25mT. The corresponding spectra of $|\Delta L_{21}|$ and $|\Delta L_{11}|$ for $d=20 \mu$m are shown in Fig. 4a in the main text. The solid lines represent the theoretical fitting by the exponential decay function $2|\Delta L_{21}|/|\Delta L_{11}| = \exp (-d/l)$. {\bf b}, Temperature dependence of damping parameter $\alpha$, obtained from the device with $d=40 \mu$m.}
\label{GapDep}
\end{center}
\end{figure*}

In general, the ratio between the self-inductance $|\Delta L_{11}|$ and mutual inductance $|\Delta L_{21}|$ provides the decay rate of spin excitation amplitude during the propagation for the gap distance $d$ between two CPWs used for the excitation and detection\cite{AESWS_doppler_B}. In Fig. \ref{GapDep}a, the $2|\Delta L_{21}|/|\Delta L_{11}|$ value (deduced from the similar experimental data sets as shown in Fig. 4a in the main text) is plotted for various excitation modes in the SkX state as a function of $d$. It roughly follows the expected $2|\Delta L_{21}|/|\Delta L_{11}| = \exp (-d/l)$ relationship\cite{AESWS_doppler_B}, where the slight deviation from this curve may be ascribed to the device dependence of the unintentional air gap between the CPWs and Cu$_2$OSeO$_3$ sample. 

Here, the decay length $l$ is related with the damping parameter $\alpha$ in form of  $l=v_g/(2\pi \nu_0 \alpha)$\cite{AESWS_doppler_B}. In Fig. \ref{GapDep}b, temperature dependence of $\alpha$ value for various excitation modes, experimentally deduced for the device with $d=40\mu$m, is plotted. The corresponding data for the device with $d=20\mu$m are also shown in Fig. 4e in the main text. In both cases, the SkX phase always hosts slightly larger, but less than twice of, $\alpha$ value as compared to the ferromagnetic phase. The order of $\alpha$ values for these two devices also roughly agree with each other. These results confirm the overall reliability of observed mode dependence of $\alpha$ and $l$. (Note that the high temperature data for the $d=40\mu$m device become less reliable than the $d=20\mu$m device, because the $2|\Delta L_{21}|/|\Delta L_{11}|$ value for the former one is  more suppressed due to the longer distance of propagation.)

\subsection{Dependence on the geometrical configuration}

\begin{figure}[b]
\begin{center}
\includegraphics*[width=14cm]{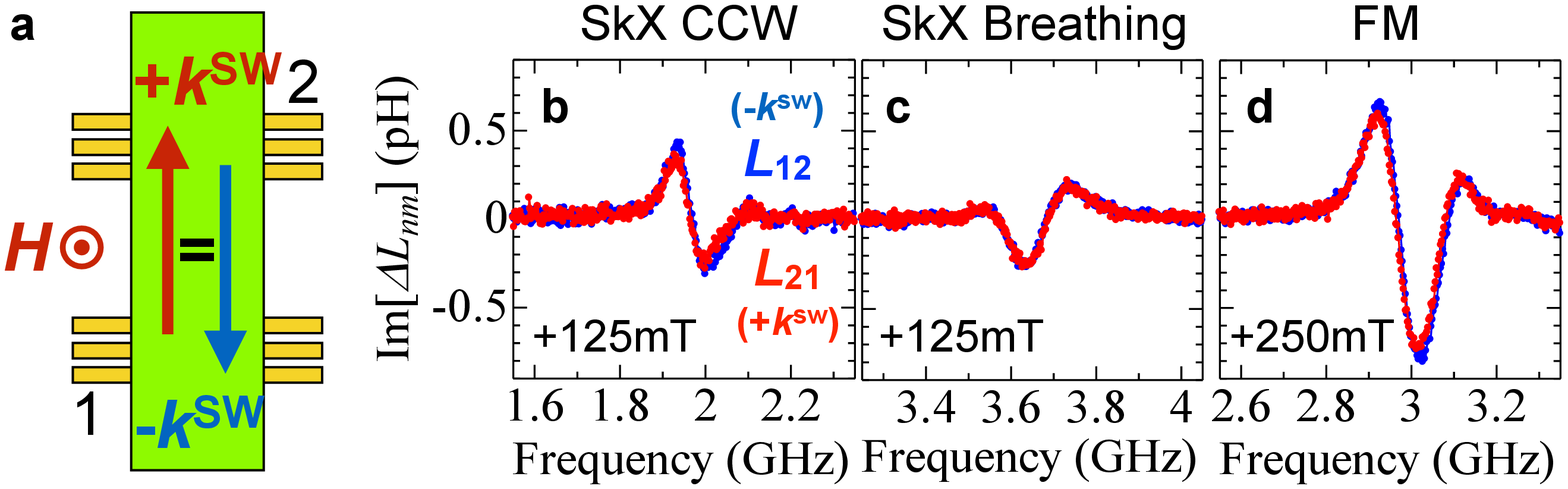}
\caption{{\bf a}, Schematic illustration of the measurement configuration for the $H \perp k^{\rm SW}$ setup. {\bf b-d}, The spectra of mutual inductance $\Delta L_{21}$ and $\Delta L_{12}$, which represent the propagation character of spin excitation with the wave vector $+k^{\rm SW}$ and $-k^{\rm SW}$, respectively. All the data were measured at 30 K under the configuration shown in {\bf a}. Here, {\bf b} and {\bf c} represent the CCW and breathing modes in the SkX state at +125 mT, respectively, and {\bf d} indicates the resonance mode in the collinear ferromagnetic state at +250 mT. Note that the $H$-value required for the stabilization of the SkX phase is different between the $H \perp k^{\rm SW}$ and $H\parallel k^{\rm SW}$ configurations, due to the demagnetization effect.} 
\label{GeometryDep}
\end{center}
\end{figure}

In the main text, we discussed the results for the $H \parallel k^{\rm SW}$ configuration (Fig. 2d), where the spin excitation propagates parallel to the skyrmion strings in the SkX state. For the $H \perp k^{\rm SW}$ configuration (Fig. \ref{GeometryDep}a), on the other hand, we can investigate the character of spin excitation propagating normal to the skyrmion strings. Figures \ref{GeometryDep}b and c indicate the spectra of mutual inductance $\Delta L_{21}$ and $\Delta L_{12}$ (representing the propagation character of spin excitation with the wavevector $+k^{\rm SW}$ and $-k^{\rm SW}$) measured with the $H \perp k^{\rm SW}$ configuration for the CCW and breathing modes in the SkX state (Note that the signal of the CW mode is too weak to be detected here). The corresponding data measured in the ferromagnetic state is also plotted in Fig. \ref{GeometryDep}d. Unlike the case of $H \parallel k^{\rm SW}$, nonreciprocal propagation behavior or associated frequency shift $\Delta \nu$ between $\pm k^{\rm SW}$ have not been observed for any mode in the $H \perp k^{\rm SW}$ configuration. Such an absence of nonreciprocity is consistent with symmetry analysis; For the present $H \perp k^{\rm SW}$ configuration (Fig. \ref{GeometryDep}a), the two-fold rotational symmetry around $H$ is sustained, which requires the equivalent nature of $+k^{\rm SW}$ and $-k^{\rm SW}$. The above results demonstrate that the appearance or absence of nonreciprocity is strongly dependent on the geometrical relationship between the directions of $k^{\rm SW}$-vector and skyrmion strings.

\section{Theoretical calculations}
\renewcommand\vec{\mathbf}

We present details of our theoretical analysis supporting the conclusion in the main text. The computations are performed in the framework of linear spin wave theory for the skyrmion crystal following Refs.~\onlinecite{B20_FMR} and \onlinecite{WaiznerPhD}. We present  in section \ref{subsec:TheoryBasics}  the energy functional  for chiral magnets and the wave equation for the spin excitations. We also specify the parameters entering the theory and explain the connection to experiment. In sections \ref{subsec:FPphase} and \ref{subsec:conical} we discuss the field-polarized and conical phase, respectively. The numerical solution for the skyrmion crystal phase is explained in section \ref{subsec:SkX1}. The section \ref{subsec:SkX2} focusses on the non-reciprocity of spin waves in the skyrmion crystal phase. It is demonstrated that it derives both from the DM interaction and the stray field energy with the main results given in Eqs.~\eqref{vinftyDMI} and \eqref{vinftydip}.

\subsection{Linear spin wave theory for cubic chiral magnets}
\label{subsec:TheoryBasics}

\subsubsection{Free energy and equation of motion}

The magnetic properties of the cubic chiral magnets deep within the order magnetic phases are well described in terms of a unit vector $\vec m({\bf r})$ representing the direction of the magnetization. It is governed by the free energy functional $\mathcal{F} = \mathcal{F}_0 + \mathcal{F}_{\rm dip}$ with 
\begin{align} \label{F0}
\mathcal{F}_0 &= \int d{\bf r} \Big[ \frac{J}{2} (\nabla_j \vec m_i)^2 + D \vec m (\nabla \times \vec m) - \mu_0 M_s H \vec m_z + \lambda (\vec m^2 - 1)^2\Big]
\end{align}
where $J$ is the exchange interaction, $D$ is the DM interaction, $M_s$ is the saturated magnetization, $\mu_0$ is the magnetic constant, and $H$ is the magnetic field applied along the $z$-axis. In the following, we assume a right-handed magnetic system with positive $D>0$. It is convenient to impose the condition $\vec m^2 ({\bf r}) = 1$ approximately with the help of a 'soft-spin' implementation represented by the last term in Eq.~\eqref{F0}. In the following, we use  a fixed $\lambda = 160 000$ so that the length $|\vec m({\bf r})|$ varies, for example, by less than half a per mille within the skyrmion crystal phase.

The stray field energy reads
\begin{align} \label{Fdip}
\mathcal{F}_{\rm dip} &= \int \frac{d{\bf k}}{(2\pi)^3} \frac{1}{2} \vec m_i(-{\bf k}) \chi^{-1}_{{\rm dip},ij}({\bf k}) \vec m_j({\bf k})
\end{align}
with the Fourier transform $\vec m({\bf k}) = \int d{\bf r} e^{- i{\bf k}{\bf r}} \vec m({\bf r})$. For wavevectors much larger than the inverse linear size of the sample $|{\bf k}| \gg 1/L$, the susceptibility is given by $\chi^{-1}_{{\rm dip},ij}({\bf k}) = \mu_0 M_s^2 \frac{{\bf k}_i {\bf k}_j}{{\bf k}^2}$. For zero wavevector it is determined by the demagnetization factors $\chi^{-1}_{{\rm dip},ij}(0) = \mu_0 M_s^2 N_{ij}$ with $N_{ij} =$ diag$(N_x,N_y,N_z)$ for an ellipsoidal sample. For intermediate wavevectors 
$|{\bf k}| L \lesssim 1$, the system is in the magnetostatic limit where the magnetic properties depend on the details of the sample size.

The equation of motion that governs the magnetization dynamics is given by 
\begin{align} \label{EoM}
\partial_t \vec m = - \gamma \vec m \times \vec B_{\rm eff}
\end{align}
with the effective field $\vec B_{\rm eff} = -\frac{1}{M_s} \frac{\delta \mathcal{F}}{\delta \vec m}$ and the gyromagnetic ratio $\gamma = g \mu_B/\hbar$. The magnetic phases are identified in equilibrium by a vanishing effective magnetic field $\vec B_{\rm eff}|_{\vec m_0} = 0$ when evaluated with the equilibrium magnetization $\vec m_0({\bf r})$. In first order in the deviation $\delta \vec m({\bf r},t) = \vec m({\bf r},t) - \vec m_0({\bf r})$ the field $\vec B_{{\rm eff},i}({\bf r},t) =  -\frac{1}{M_s}  \int d{\bf r'} \chi_{ij}^{-1}({\bf r},{\bf r'}) \delta \vec m_j({\bf r'},t)$ is then given in terms of the susceptibility $\chi_{ij}^{-1}({\bf r},{\bf r'}) = \delta^2 \mathcal{F}/(\delta \vec m_i({\bf r})\delta \vec m_j({\bf r'}))|_{\vec m_0}$ that is to be evaluated with the equilibrium magnetization $\vec m_0({\bf r})$. The linear spin wave theory is obtain by expanding the equation of motion \eqref{EoM} in first order in $\delta \vec m$,
\begin{align} \label{SpinwaveEq}
\partial_t \delta \vec m({\bf r},t) = \frac{\gamma}{M_s} \vec m_0({\bf r}) \times \int d{\bf r'} \chi^{-1}({\bf r},{\bf r'}) \delta \vec m({\bf r'},t)
 \end{align}
The spin wave spectrum and its eigenvectors are obtained by solving the spin wave equation \eqref{SpinwaveEq}.

\subsubsection{Parameters entering the theory}
\label{subsubsec:Parameters}

Important parameters are the wavevector $Q = D/J$, the internal critical field $\mu_0 H^{\rm int}_{c2} = D^2/(J M_s)$ separating the conical from the field-polarized phase, and the susceptibility in the conical phase $\chi_{\rm con}^{\rm int} = \frac{\mu_0 M_s^2}{J Q^2}$. For Cu$_2$OSeO$_3$ we use $Q = 2\pi/\lambda$ with $\lambda = 60$ nm and 
$\chi_{\rm con}^{\rm int} = 1.76$ \cite{B20_FMR}. The critical field $H^{\rm int}_{c2}$ is temperature dependent and for $T = 25$ K we have $\mu_0 H^{\rm int}_{c2} \approx 0.07$~T. After measuring length in units of $1/Q$ and energy in units of $g \mu_B \mu_0 H^{\rm int}_{c2}$ with $g$ factor $g \approx 2.1$, the theory then only depends on the parameters given by the demagnetization factors of the sample and the ratio $H/H_{c2}$ measuring the strength of the applied magnetic field with $H_{c2} = H^{\rm int}_{c2}(1+N_z \chi_{\rm con}^{\rm int})$. In the present study, the investigated plate-shape sample of Cu$_2$OSeO$_3$ can be approximately characterized by the demagnetization factors: 
\begin{align} \label{DemagFactors}
N_x = 0.879,\quad N_y = 0.105, \quad N_z = 0.016,
\end{align}
where the magnetic field is applied along the $z$-axis. In the following discussion, we will also use the abbreviation $\mathcal{D} = g \mu_B \mu_0 H^{\rm int}_{c2}/Q^2$ for the stiffness.

\subsubsection{Limits of the spin wave dispersion and experimental quantities}

In the following, we discuss volume spin waves and their dispersion $\omega({\bf k}) = 2\pi \nu({\bf k})$. 
We distinguish between the bulk spin wave dispersion for wavevectors $|{\bf k}| L \gg 1$ and the magnetostatic limit 
of the spin wave dispersion for wavevectors $|{\bf k}| L \lesssim 1$ with the linear size $L$ of the sample. It is important to note that the limits of small wavevectors ${\bf k} \to 0$ and a large bulk sample $L \to \infty$ do not commute. The precise form of the dispersion  in the magnetostatic limit depends on the details of the sample shape and the boundary conditions, and we will not attempt to provide a discussion of the dispersion in this regime. We will limit ourselves to a discussion of the bulk dispersion for $|{\bf k}| L \gg 1$ and the uniform resonance frequency at zero wavevector, $\nu(0)$. In particular, we focus on the bulk spin wave dispersion for wavevectors longitudinal to the applied magnetic field ${\bf H} = H \hat z$ that is non-reciprocal $\omega(k_z) \neq \omega(-k_z)$ in chiral magnets and gives access to the following quantities
\begin{align} \label{Limits}
\nu_{\infty} \equiv \lim_{k_z \to 0} \lim_{L \to \infty} \frac{\omega(k_z)}{2\pi},\quad 
v_{\infty} \equiv \lim_{k_z \to 0} \lim_{L \to \infty} \partial_{k_z} \omega(k_z).
\end{align}

The limit $\nu_\infty$ and the uniform resonance frequency $\nu(0)$ differ, and the dispersion $\nu(k_z)$ interpolates between these two values in the magnetostatic limit $|k_z| L \lesssim 1$ as sketched, e.g. in Fig.~2j of the main text. Magnetostatic modes with $\nu(0) > \nu_\infty$ and $\nu(0) < \nu_\infty$ are known, respectively, as backward (BVMSW) and forward (FVMSW) volume magnetostatic spin wave modes. The two values $\nu_\infty$ and $\nu(0)$ are shown in Fig.~3d of the main text as dashed and solid lines, respectively. The experiment is performed in the magnetostatic limit (as $|k^{\rm SW}| b \sim 1$ in our setup) so that the recorded frequency $\nu(k_z)$, also denoted by $\nu_0$ in the main text, is located within the frequency range enclosed by $\nu_\infty$ and $\nu(0)$. As the interpolation between these two values occurs on the scale of a wavevector given by the inverse thickness of the sample $1/b$, we can crudely estimate the experimentally measured group velocity by $v_g \approx 2\pi (\nu(0) - \nu_{\infty}) b$. 

The velocity $v_\infty$ in Eq.~\eqref{Limits} is a measure of the non-reciprocity of the bulk spectrum. We assume that it also determines the non-reciprocity in the magnetostatic limit (with the conical phase being an exception, see below) so that we approximate $\Delta \nu(k_z) = \nu(k_z) - \nu(-k_z) \approx 2 v_\infty k_z/(2\pi)$. 

We summarize the relation between the measured quantities $\nu_0 = \nu(k_z), \Delta \nu(k_z)$ and $v_g(k_z)$ shown in Fig.~3 of the main text and the theoretically accessible parameters $\nu(0), \nu_{\infty}$ and $v_\infty$, 
\begin{align}
\nu_0 &\in \{\min \{\nu_{\infty}, \nu(0)\}, \max  \{\nu_{\infty}, \nu(0)\}  \},\\
\label{Nonreciprocity}
\Delta \nu &= \nu(k_z) - \nu(-k_z) \approx \frac{2 v_\infty k_z}{2\pi},\\
v_g &\approx 2\pi (\nu(0) - \nu_{\infty}) b.
\end{align}
Here, the wavevector $k_z$ denoted by $k^{\rm SW}$ in the main text is assumed to be in the magnetostatic limit $|k^{\rm SW}| b \sim 1$ with the width $b$ of the sample. These approximations allow for a comparison with theory without the computation of the full dispersion in the magnetostatic limit, which is a formidable task and requires much more effort. Note that all parameters entering the calculation, see section \ref{subsubsec:Parameters}, are known from independent measurements so that the theory provides quantitative parameter-free predictions for $\nu(0), \nu_{\infty}$ and $v_\infty$.

In the next sections, we discuss the theoretical values $\nu(0), \nu_{\infty}$ and $v_\infty$ for the various magnetic phases.

\subsection{Field-polarized phase}
\label{subsec:FPphase}

In the field-polarized (FP) phase at $H>H_{c2}$, the magnetic ground state within the bulk of the sample is polarized along the applied field, $\vec m_0 = \hat H = \hat z$. The bulk spin wave spectrum for a wavevector aligned with the magnetic field is given by \cite{Garst2017}
\begin{align}
\hbar \omega(k_z) = 2 \mathcal{D} Q k_z + \mathcal{D} k_z^2 + g \mu_B \mu_0 H_{\rm int}
\end{align}
with the internal field $H_{\rm int} = H - N_z M_s$. We obtain for the quantities of Eqs.~\eqref{Limits}
\begin{align}
\nu_{{\rm FP},\infty} = \frac{g \mu_B \mu_0 H_{\rm int}}{2\pi \hbar},\quad 
v_{{\rm FP},\infty} = \frac{2 \mathcal{D} Q}{\hbar} 
= 2 \frac{g \mu_B \mu_0 H^{\rm int}_{c2}}{\hbar Q}.
\end{align}
The excitation energy at strictly zero wavevector ${\bf k} = 0$ is given by the Kittel formula
\begin{align}
\nu_{{\rm FP}}(0) = \frac{g \mu_B \mu_0}{2\pi \hbar} \sqrt{(H - (N_z - N_x) M_s) (H - (N_z - N_y) M_s)}.
\end{align}
In the geometry \eqref{DemagFactors} the Kittel frequency is larger than $\nu_{{\rm FP},\infty}$ so that the spin wave is a BVMSW with a dynamic magnetization that oscillates within the plane perpendicular to the wavevector ${\bf k} = k_z \hat z$.

\subsection{Conical phase}
\label{subsec:conical}

The magnetization of the conical phase for $H<H_{c2}$ is given by $\vec m_0(z) = (\sin\theta \cos Q z, \sin\theta \sin Q z, \cos \theta)$ with the cone angle $\theta$ that obeys $M_s \cos \theta = \chi_{\rm con}^{\rm int} H_{\rm int}$. The periodicity of the magnetization leads to Bragg scattering of spin waves and a magnon band structure \cite{Garst2017}. The 
bulk spin wave spectrum for ${\bf k} = k_z \hat z$ in the extended zone scheme is given by 
\begin{align} \label{BulkDispConical}
\hbar \omega(k_z) = \mathcal{D} |k_z| \sqrt{k_z^2 + (1+\chi_{\rm con}^{\rm int}) Q^2 \left(1-\left(\frac{H_{\rm int}}{H^{\rm int}_{c2}}\right)^2\right)}.
\end{align}
There are two magnetic resonances in the conical phase denoted by $+Q$ and $-Q$ whose limit $\nu_{\infty}$ for small wavevectors $k_z \to 0$ is degenerate in the repeated zone scheme, and it is obtained by the taking the limit $k_z \to \pm Q$ of Eq.~\eqref{BulkDispConical}, 
\begin{align}
\nu_{\pm Q,\infty} = \frac{\omega(\pm Q)}{2\pi} = 
\frac{g \mu_B \mu_0 H^{\rm int}_{c2}}{2\pi\hbar} \sqrt{1 + (1+\chi_{\rm con}^{\rm int})  \left(1-\left(\frac{H_{\rm int}}{H^{\rm int}_{c2}}\right)^2\right)}.
\end{align}

The computation of the group velocity of the bulk spectrum $v_\infty$ is tricky due to the degeneracy $\nu_{\pm Q,\infty}$. Considering the derivative of Eq.~\eqref{BulkDispConical} in the limit $k_z \to \pm Q$, one finds a finite velocity with $v_{+Q,\infty} = - v_{-Q,\infty}$, i.e., the dispersion of the $\pm Q$ modes cross at $\nu_{\pm Q,\infty}$. This crossing becomes however an avoided crossing in the magnetostatic limit where the stray field lifts the degeneracy. For this reason, we argue that in the conical phase the equation \eqref{Nonreciprocity} is not applicable, and instead the non-reciprocity is practically vanishing $\Delta \nu \approx 0$. 

We note however that higher-order gradient corrections to the theory of Eq.~\eqref{F0} can shift the crossing point of the bulk spectrum at energies $\nu_{\pm Q,\infty}$ away from $k_z = 0$ in the repeated zone scheme, see the discussion in Refs.~\cite{Kugler2015,Weber2017}. This leads to a small non-reciprocity even in the conical phase that decreases with the applied magnetic field. This effect is neglected in the theoretical figure of Fig.~3 in the main text. 

The uniform resonance frequencies of the two modes $\nu_{\pm Q}(0)$ are known in closed form and given in Refs.~\cite{B20_FMR,Garst2017} so that we do not repeat them here. Both modes are BVMSW, $\nu_{\pm Q}(0) > \nu_{\pm Q,\infty}$, as their mean magnetization oscillates within the plane perpendicular to the wavevector ${\bf k} = k_z \hat z$. However, the spectral weight of the $-Q$ mode is much smaller in the investigated field range \cite{Garst2017} so that only the $+Q$ mode is detected in the experiment. Consequently, only the theoretical results for the $+Q$ mode is presented in Fig.~3 of the main text.

\subsection{Skyrmion crystal phase -- numerical solution of the spin wave spectrum}
\label{subsec:SkX1}

\subsubsection{Variational Ansatz and spin wave equation}

In order to obtain the magnetization of the skyrmion crystal, we use the variational Ansatz 
\begin{align} \label{SkXAnsatz}
\vec m_0({\bf r}) = \sum_{{\bf G}_\perp \in L_R} \vec m_0({\bf G}_\perp) e^{i {\bf G}_\perp {\bf r}} 
\end{align}
with the Fourier components $\vec m_0({\bf G}_\perp)$ where the vectors ${\bf G}_\perp$ belong to the two-dimensional triangular reciprocal lattice $L_R$ that is perpendicular to the applied magnetic field, ${\bf G}_\perp \hat z = 0$. In practice, the reciprocal lattice is restricted to a finite number of primitive unit cells of the reciprocal lattice and the symmetries of the skyrmion crystal are  exploited in order to reduce the amount of variational parameters $\vec m_0({\bf G}_\perp)$, for details see Ref.~\cite{WaiznerPhD}. First, the free energy is minimized with the Ansatz \eqref{SkXAnsatz} and, in a second step, the spin wave equation \eqref{SpinwaveEq} is solved. 

With the help of the Fourier transforms $\delta \vec m({\bf r}, t) = \int \frac{d{\bf q}}{(2\pi)^3} \frac{d\omega}{2\pi} e^{- i \omega t + i {\bf q r}} \delta \vec m({\bf q},\omega)$ and $\chi^{-1}({\bf r},{\bf r'}) = \int \frac{d{\bf q}}{(2\pi)^3}\frac{d{\bf q'}}{(2\pi)^3} e^{i {\bf q r}+i {\bf q' r'}} \chi^{-1}({\bf q},{\bf q'})$ this wave equation can be expressed as
\begin{align} 
-i \omega \delta \vec m({\bf q},\omega) = \frac{\gamma}{M_s} \sum_{{\bf G}_\perp \in L_R}  \int \frac{d{\bf q'}}{(2\pi)^3} \vec m_0({\bf G}_\perp) \times \left( \chi^{-1}({\bf q} - {\bf G}_\perp,-{\bf q'}) \delta \vec m({\bf q'},\omega) \right).
\end{align}
Decomposing the wavevectors ${\bf q} = {\bf K}_\perp + {\bf k}$ and ${\bf q'} = {\bf K'}_\perp + {\bf k'}$ into reciprocal lattice vectors, ${\bf K}_\perp$ and ${\bf K'}_\perp$, and wavevectors ${\bf k}$ and ${\bf k'}$ whose components perpendicular to the $z$-axis belong to the first Brillouin zone, ${\bf k}_\perp, {\bf k'}_\perp \in $ 1.BZ, we can exploit that the susceptibility is diagonal in  wavevectors but only up to reciprocal lattice vectors, $\chi^{-1}_{ij}({\bf K}_\perp + {\bf k} - {\bf G}_\perp,-{\bf K'}_\perp - {\bf k'}) = 
\chi^{-1}_{ij}({\bf K}_\perp  - {\bf G}_\perp,-{\bf K'}_\perp; {\bf k})(2\pi)^3 \delta ({\bf k} - {\bf k'})$, so that the wave equation simplifies to
\begin{align} \label{SWeq1}
\omega \delta \vec m({\bf K}_\perp+ {\bf k},\omega) = \sum_{{\bf K'}_\perp \in L_R} \mathcal{W}({\bf K}_\perp,{\bf K'}_\perp;{\bf k})
\delta \vec m({\bf K'}_\perp +  {\bf k},\omega)
\end{align}
with the matrix 
\begin{align} \label{Matrix}
\mathcal{W}_{nm}({\bf K}_\perp,{\bf K'}_\perp;{\bf k}) = 
i\frac{\gamma}{M_s} \sum_{{\bf G}_\perp \in L_R}  
\epsilon_{n \ell j}
\vec m_{0,\ell}({\bf G}_\perp) 
\chi^{-1}_{jm}({\bf K}_\perp  - {\bf G}_\perp,-{\bf K'}_\perp; {\bf k}) .
\end{align}
The solution for a given frequency $\omega$ and wavevector ${\bf k}$ yields the dispersion $\omega({\bf k})$ and the eigenvectors 
$\delta \vec m({\bf K}_\perp+ {\bf k},\omega)$.

\subsubsection{Numerical solution for the spin wave spectrum}

\begin{figure}
\includegraphics[width=0.5\linewidth]{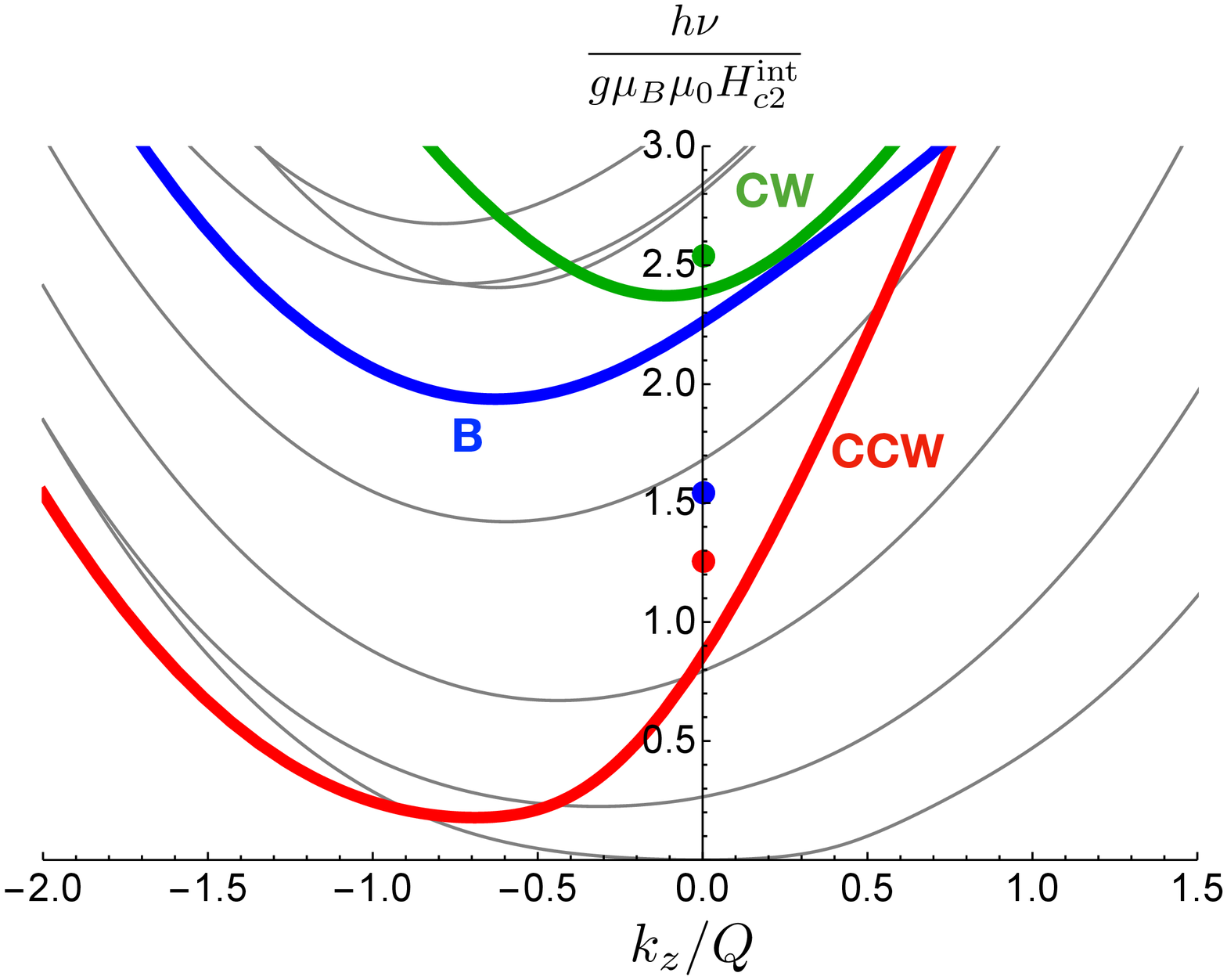}
\caption{\label{fig:SkXspectrum} The bulk spin wave spectrum $\nu(k_z) = \omega(k_z)/(2\pi)$ for the skyrmion crystal in Cu$_2$OSeO$_3$ numerically evaluated for a magnetic field $H/H_{c2} = 0.4$ as a function of wavevector $k_z$ parallel to the applied magnetic field. There are various modes (grey) but only the CCW (red), B (blue) and CW (green) modes possess a global dynamic magnetization. The uniform resonance frequency $\nu(0)$ at $k_z = 0$ of these modes for the geometry of Eq.~\eqref{DemagFactors} is also indicated by the dots. 
}
\end{figure}

The numerical solution for the bulk spin wave spectrum for wavevectors perpendicular to the applied field, $\omega({\bf k}_\perp)$, was presented, e.g., in Ref.~\cite{Garst2017}. Here, we discuss the spectrum for wavevectors along the field $\omega(k_z)$. In Fig.~\ref{fig:SkXspectrum} the spectrum numerically evaluated for a magnetic field $H = 0.4 H_{c2}$ is shown. There are various modes but only three of them possess a global dynamic magnetization, i.e., an oscillating magnetic dipole moment on average: the counterclockwise (CCW), the breathing (B) and the clockwise (CW) mode indicated by the colored lines. Their dispersion is reproduced in Fig.~2i of the main text. The uniform resonance frequency $\nu(0)$ of these modes  at zero wavevector is represented by the colored dots for the geometry of Eq.~\eqref{DemagFactors}. 
We can conclude that the CCW and the CW modes are BVMSW modes as their $\nu_\infty < \nu(0)$ whereas the breathing mode is a FVMSW mode with $\nu_\infty > \nu(0)$. This is consistent with the observation that their global dynamic magnetization oscillates within the plane perpendicular to the wavevector ${\bf k} \parallel \vec H$ for the CCW and CW modes (in a counterclockwise and clockwise manner, respectively), and it is linearly polarized along the wavevector ${\bf k} \parallel \vec H$ for the B mode. The numerically computed values of $\nu(0)$ and $\nu_\infty$ for the three modes at various magnetic fields are shown in Fig.~3d of the main text by the solid and dashed lines, respectively. 

Moreover, the slope $v_\infty$ of the spectra in Fig.~\ref{fig:SkXspectrum} close to zero wavevector, that is a measure for the non-reciprocity, is the largest for the CCW mode. In the next section, we present an analytical expression for this slope $v_\infty$ in terms of the spin wave function that elucidates the different non-reciprocities present in Fig.~\ref{fig:SkXspectrum}.

\subsection{Skyrmion crystal phase -- non-reciprocity of the bulk spin wave spectrum}
\label{subsec:SkX2}

We aim to derive an analytical expression for the slope $v_\infty$ of the bulk spin wave dispersions for the skyrmion crystal that quantifies their non-reciprocity. When we consider the limit $k_z \to 0$ in the following, it is implied that the limit $L \to \infty$ for an infinitely large sample was taken first in order to comply with the definition of Eq.~\eqref{Limits}

\subsubsection{Solution of the spin wave equation for ${\bf k}_\perp = 0$ and $k_z \to 0$} 

\begin{figure}
\includegraphics[width=\linewidth]{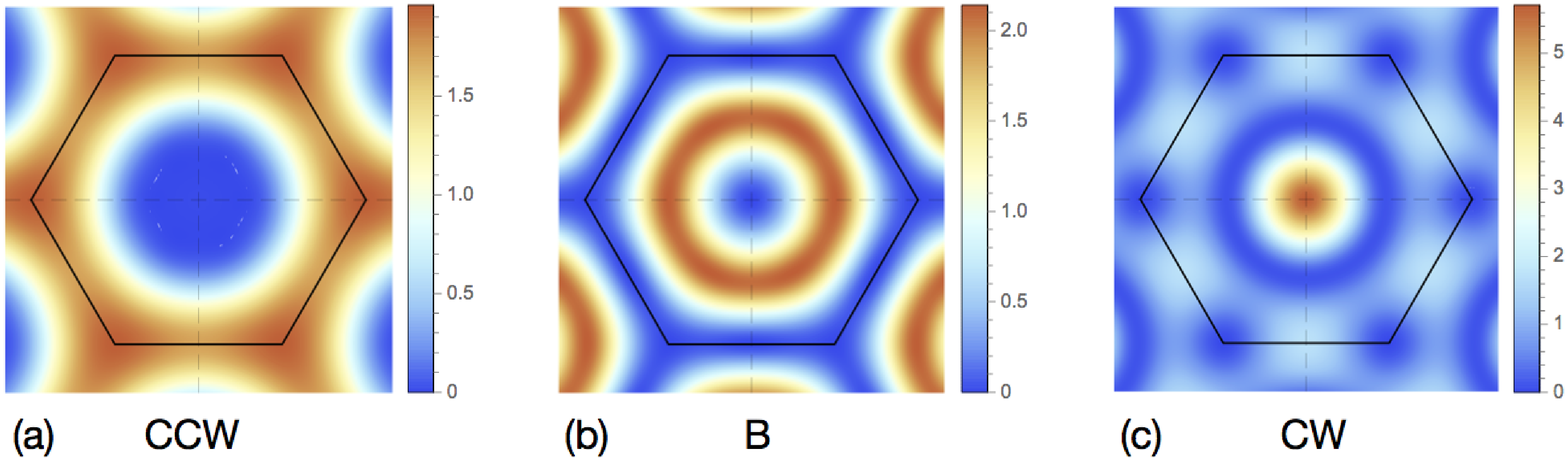}
\caption{\label{fig:norm} {\bf a-c}, Integrand $\mathcal{A}({\bf r}_\perp) = \delta \vec m^\dagger_{\alpha}({\bf r}_\perp) i 
\left(\vec m_0({\bf r}_\perp) \times \delta \vec m_{\alpha}({\bf r}_\perp)\right) $ of the normalization condition \eqref{normalization} in the two-dimensional plane perpendicular to the magnetic field for the CCW, breathing and CW mode at the magnetic field $H = 0.4 H_{c2}$. The black lines indicate the primitive unit cell of the magnetic skyrmion crystal.}
\end{figure}

In this limit the spin wave equation \eqref{SWeq1} reads
\begin{align}  \label{SWeq2}
\omega \delta \vec m({\bf K}_\perp,\omega) = \sum_{{\bf K'}_\perp \in L_R} \mathcal{W}({\bf K}_\perp,{\bf K'}_\perp;0 \hat z)
\delta \vec m({\bf K'}_\perp,\omega),
\end{align}
where $0 \hat z$ in the last argument of the matrix $\mathcal{W}$ defined in Eq.~\eqref{Matrix} is a reminder of the particular limit we are considering. The eigenvectors with eigenfrequency $\omega_\alpha$ will be denoted by $\delta \vec m_\alpha({\bf K}_\perp)$. These eigenfrequencies identify $\nu_{\infty}$, see Eq.~\eqref{Limits}, of the mode with quantum number $\alpha$. Note that the matrix $\mathcal{W}$ is non-hermitian so that the system must be solved with a Bogoliubov transformation instead of a unitary transformation. As a consequence, the eigenvectors (for positive eigenvalues) are orthonormal with respect to the scalar product
\begin{align}
\langle \delta m_{\alpha'}  | \delta  m_\beta \rangle = \sum_{{\bf G}_\perp, {\bf G'}_\perp \in L_R} 
\delta \vec m^\dagger_{\alpha'}({\bf G'}_\perp) i \left(\vec m_0({\bf G'}_\perp-{\bf G}_\perp) \times \delta\vec m_{\alpha}({\bf G}_\perp)\right) = \delta_{\alpha,\alpha'}.
\end{align}
It is convenient to express this relation in real space with the help of the Fourier transform $\delta \vec m_{\alpha}({\bf r}_\perp) = \sum_{{\bf G}_\perp \in L_R} e^{i {\bf G}_\perp {\bf r}_\perp} \delta \vec m_{\alpha}({\bf G}_\perp)$ and the standard relations
\begin{align}
\frac{1}{V_{\rm UC}} \sum_{{\bf G}_\perp \in L_R} e^{i {\bf G}_\perp {\bf r}_\perp} = \delta({\bf r}_\perp), \qquad
\int_{V_{\rm UC}} d{\bf r}_\perp e^{i {\bf G}_\perp {\bf r}_\perp} = V_{\rm UC} \delta_{{\bf G}_\perp, 0},
\end{align}
where the integral in the second equation is over the two-dimensional (primitive) unit cell of the magnetic skyrmion crystal with volume $V_{\rm UC}$. The orthogonality relation can then be expressed as 
\begin{align} \label{normalization}
\frac{1}{V_{\rm UC}} \int_{V_{\rm UC}} d{\bf r}_\perp \delta \vec m^\dagger_{\alpha'}({\bf r}_\perp) i 
\left(\vec m_0({\bf r}_\perp) \times \delta \vec m_{\alpha}({\bf r}_\perp)\right) = \delta_{\alpha,\alpha'}.
\end{align}
The quantity $i \left(\delta \vec m_{\alpha}({\bf r}_\perp) \times \delta \vec m^*_{\alpha'}({\bf r}_\perp) \right) = \mathcal{A}({\bf r}_\perp)\, \vec m_0({\bf r}_\perp)$ possesses a direction that is given by the local magnetization and a magnitude $\mathcal{A}$ that is proportional to the area enclosed by the local precession of the magnetization. The latter is normalized by the condition \eqref{normalization} so that $\frac{1}{V_{\rm UC}} \int_{V_{\rm UC}} d{\bf r}_\perp \mathcal{A}({\bf r}_\perp) = 1$ (assuming that $\vec m^2_0({\bf r}_\perp) = 1$). Examples for the density distribution $\mathcal{A}({\bf r}_\perp)$ can be found in Fig.~1k-m of the main text that are repeated in Fig.~\ref{fig:norm} with a different coloring.

\subsubsection{Analytical expression for the velocity $v_\infty$} 
In order to determine the velocity $v_\infty$ we can apply perturbation theory in the wavevector $k_z$. Consider first the contribution to the matrix $\mathcal{W}$ attributed to the DM interaction and stray field energy
\begin{align}
\mathcal{W}_{nm}({\bf K}_\perp,{\bf K'}_\perp;{\bf k}) = 
i\frac{\gamma}{M_s}   
\epsilon_{n \ell j}
\vec m_{0,\ell}({\bf K}_\perp - {\bf K'}_\perp)
\left(\chi^{-1}_{{\rm DM}, jm}({\bf K'}_\perp+{\bf k}) +\chi^{-1}_{{\rm dip}, jm}({\bf K'}_\perp+{\bf k}) \right) + \dots
\end{align}
with $\chi^{-1}_{{\rm DM}, jm}({\bf q}) = 2 D \epsilon_{jnm} i {\bf q}_n$ and $\chi^{-1}_{{\rm dip}, jm}({\bf q}) = \mu_0 M_s^2 \frac{{\bf q}_j {\bf q}_m}{{\bf q}^2}$ of Eq.~\eqref{Fdip}. Both contribute to first order in $k_z$,
\begin{align}
&\mathcal{W}^{(1)}_{nm}({\bf K}_\perp,{\bf K'}_\perp;{\bf k}) =\\\nonumber
&\qquad i\frac{\gamma}{M_s}   
\epsilon_{n \ell j}
\vec m_{0,\ell}({\bf K}_\perp - {\bf K'}_\perp)
\left(2 D \epsilon_{jzm} i + \mu_0 M_s^2 \frac{{\bf K'}_{\perp,j}\delta_{m,z} +{\bf K'}_{\perp,m}\delta_{j,z} }{{\bf K'}^2_\perp}\Big|_{{\bf K'}_\perp \neq 0} \right)  k_z.
\end{align}
Treating this correction in perturbation theory we obtain a correction to the spin wave frequency
\begin{align}
&\delta \omega_\alpha(k_z) = \langle \delta m_\alpha|\mathcal{W}^{(1)} | \delta m_\alpha\rangle = \\\nonumber
&= \sum_{{\bf G}_\perp, {\bf G'}_\perp, {\bf K'}_\perp \in L_R} 
\delta \vec m^\dagger_{\alpha}({\bf G'}_\perp) i \left(\vec m_0({\bf G'}_\perp-{\bf G}_\perp) \times 
(\mathcal{W}^{(1)}({\bf G}_\perp,{\bf K'}_\perp;{\bf k})
\delta\vec m_{\alpha}({\bf K'}_\perp))\right).
\end{align}
This identifies the slope $v_{\alpha,\infty} = \delta \omega_\alpha/k_z$ for the mode with quantum number $\alpha$. 

In the following, we  discuss separately the two contributions to $v_{\alpha,\infty}$ attributed to the DM interaction and the stray field energy. The expressions for $v_{\alpha,\infty}$ become particularly transparent when they are expressed as a spatial integral over the two-dimensional magnetic unit cell.

\paragraph{Non-reciprocity due to the DM interaction.}

\begin{figure}[b]
\includegraphics[width=\linewidth]{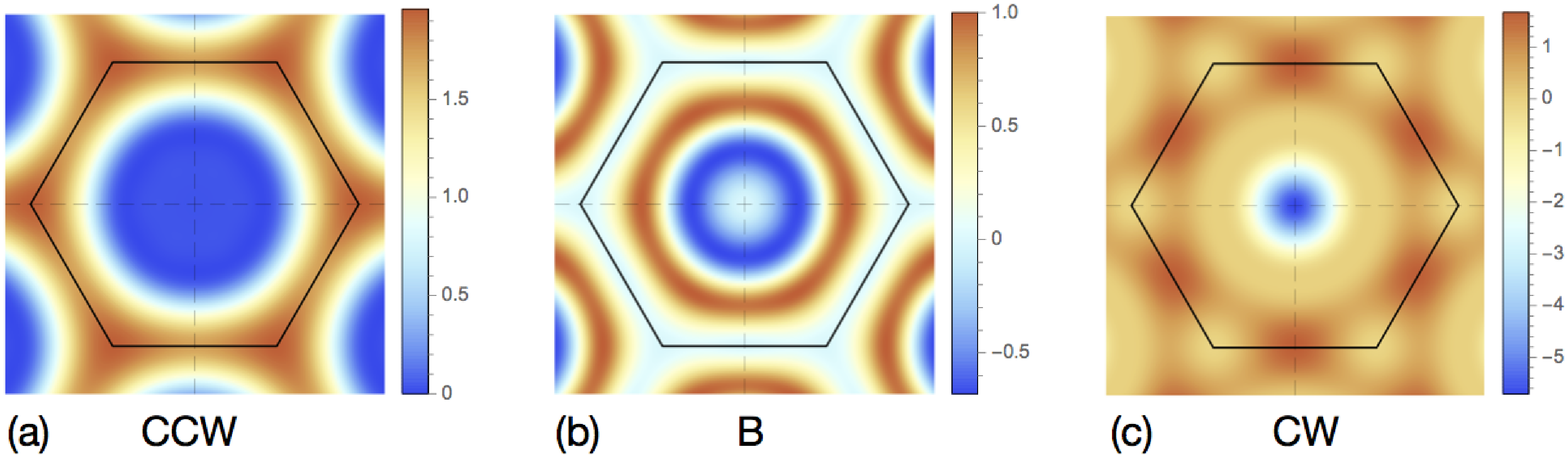}
\caption{\label{fig:DMI} {\bf a-c}, Integrand $\delta \vec m^\dagger_{\alpha}({\bf r}_\perp) i 
\left(\hat z \times \delta \vec m_{\alpha}({\bf r}_\perp)\right)$ of $v^{\rm DMI}_{\alpha,\infty}$, see Eq.~\eqref{vinftyDMI}, in the two-dimensional plane perpendicular to the magnetic field for the CCW, breathing and CW mode at the magnetic field $H = 0.4 H_{c2}$.}
\end{figure}

The contribution due to the DM interaction can be written in the form
\begin{align} \label{vinftyDMI}
\frac{v^{\rm DMI}_{\alpha,\infty}}{v_{{\rm FP},\infty}} = \frac{1}{V_{\rm UC}} \int_{V_{\rm UC}} d{\bf r}_\perp 
\delta \vec m^\dagger_{\alpha}({\bf r}_\perp) i 
\left(\hat z \times \delta \vec m_{\alpha}({\bf r}_\perp)\right) 
\end{align}
where $v_{{\rm FP},\infty} 
= 2 \frac{g \mu_B \mu_0 H^{\rm int}_{c2}}{\hbar Q}$ is the velocity of the field-polarized phase.  In the derivation we used that $\delta \vec m_{\alpha}({\bf r}_\perp) \vec m_0({\bf r}_\perp) = 0$ and $\vec m^2_0({\bf r}_\perp) = 1$. Only regions in space contribute to the integral of Eq.~\eqref{vinftyDMI} where the spin wavefunction $\delta \vec m_{\alpha}({\bf r}_\perp)$ are in-plane, i.e., orthogonal to $\hat z$. Examples for the distribution of the integrand are shown in Fig.~\ref{fig:DMI}.

As discussed in the main text, the integrand can be written, $\delta \vec m^\dagger_{\alpha}({\bf r}_\perp) i 
 \left(\hat z \times \delta \vec m_{\alpha}({\bf r}_\perp)\right) = \mathcal{A}({\bf r}_\perp) \vec m_{0,z}({\bf r}_\perp)$, as a product of the normalized density $\mathcal{A}({\bf r}_\perp) > 0$ and $\vec m_{0,z}({\bf r}_\perp)$ that varies between $-1$ and $1$. 
This implies 
\begin{align}
\frac{v^{\rm DMI}_{\alpha,\infty}}{v_{{\rm FP},\infty}} \leq \frac{1}{V_{\rm UC}} \int_{V_{\rm UC}} d{\bf r}_\perp 
 \mathcal{A}({\bf r}_\perp) = 1
\end{align}
 yielding the upper bound $v_{{\rm FP},\infty}$ for the DM contribution $v^{\rm DMI}_{\alpha,\infty}$. We conclude that in the 
 skyrmion crystal phase the non-reciprocity of spin waves attributed to the DM interaction cannot exceed the one of the field-polarized phase.

 \paragraph{Non-reciprocity due to the stray field energy.}

The additional contribution attributed to the stray field energy can be expressed in the form
\begin{align} \label{vinftydip}
\frac{v^{\rm dip}_{\alpha,\infty}}{v_{{\rm FP},\infty}} = \chi_{\rm con}^{\rm int} \frac{1}{V_{\rm UC}} \int_{V_{\rm UC}} d{\bf r}_\perp 
{\rm Im}\{\phi_{\alpha}^*({\bf r}_\perp) \delta \vec m_{\alpha,z}({\bf r}_\perp) \}
\end{align}
where we introduced the auxiliary dimensionless function
\begin{align}
\phi_{\alpha}({\bf r}_\perp) 
=  -i \sum_{ {\bf G}_\perp \in L_R; {\bf G}_\perp\neq 0}   
Q \frac{({\bf G}_{\perp}\delta \vec m_{\alpha}({\bf G}_\perp) )}{{\bf G}_\perp^2}  
  e^{i {\bf G}_\perp {\bf r}_\perp }.
\end{align}

\begin{figure}
\includegraphics[width=\linewidth]{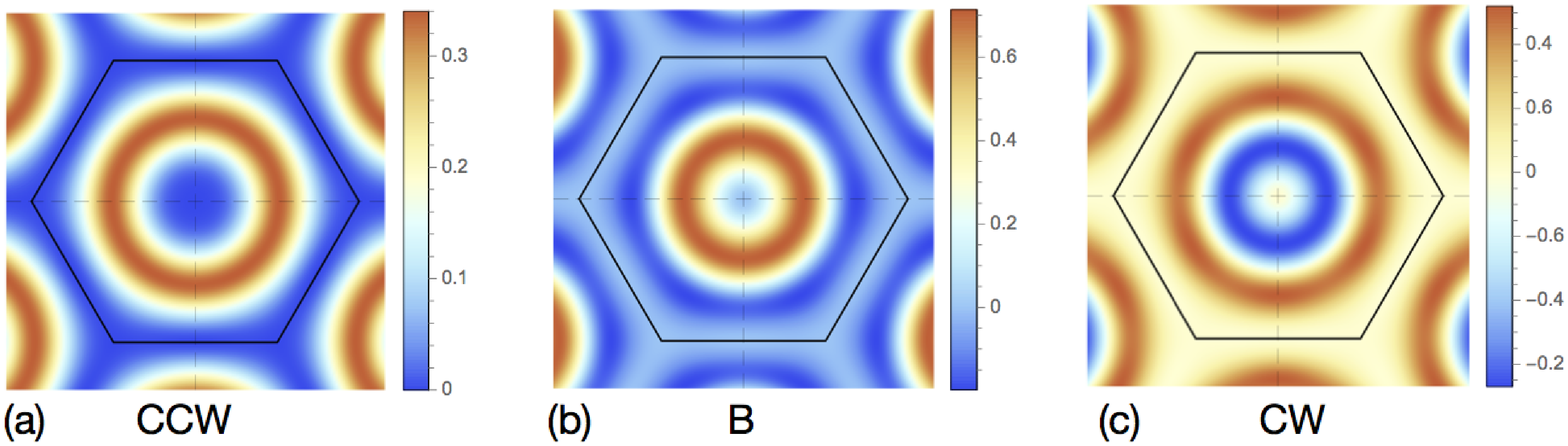}
\caption{\label{fig:dip} {\bf a-c}, Integrand ${\rm Im}\{\phi_{\alpha}^*({\bf r}_\perp) \delta \vec m_{\alpha,z}({\bf r}_\perp) \}$ of $v^{\rm dip}_{\alpha,\infty}$, see Eq.~\eqref{vinftydip}, in the two-dimensional plane perpendicular to the magnetic field for the CCW, breathing and CW mode at $H = 0.4 H_{c2}$.}
\end{figure}

In order to understand the meaning of $\phi_{\alpha}$ consider the stray field in momentum space $\vec H_{{\rm dip},\alpha}({\bf q}) $ that is generated by the spinwave function
\begin{align}
\vec H_{{\rm dip},\alpha}({\bf q}) =  - M_s {\bf q} \frac{({\bf q} \delta \vec m_{\alpha}({\bf q} ) )}{{\bf q}^2}  
\end{align}
with the saturated magnetization $M_s$. For ${\bf q}={\bf G}_\perp +{\bf k}$ at ${\bf k}_\perp = 0$ in the limit $k_z \to 0$, the corresponding field in real space reads (that is still complex as the spin wavefunction is complex)
\begin{align}
\vec H_{{\rm dip},\alpha}({\bf r}_\perp) = - M_s \sum_{ {\bf G}_\perp \in L_R; {\bf G}_\perp\neq 0 } 
{\bf G}_\perp \frac{({\bf G}_\perp \delta \vec m_{\alpha}({\bf G}_\perp) )}{{\bf G}_\perp^2} e^{i {\bf G}_\perp {\bf r}_\perp } 
- M_s \hat z \delta \vec m_{\alpha,z}(k_z \hat z)|_{k_z \to 0}
\end{align}
The last term derives from the small ${\bf q}$ limit and is spatially constant. The contribution to the stray field, that varies in space, is attributed to the first term. It turns out that $\phi_{\alpha}$ is the (dimensionless) magnetic potential that just generates this spatially dependent part of the stray field,
\begin{align}
- \frac{1}{Q}\vec \nabla_\perp \phi_\alpha({\bf r}_\perp) = - M_s \sum_{ {\bf G}_\perp \in L_R; {\bf G}_\perp\neq 0 } 
{\bf G}_\perp \frac{({\bf G}_\perp \delta \vec m_{\alpha}({\bf G}_\perp) )}{{\bf G}_\perp^2} e^{i {\bf G}_\perp {\bf r}_\perp }
 = \frac{1}{M_s} \vec H_{\rm dip}({\bf r}_\perp) +
\hat z \delta \vec m_{\alpha,z}(k_z \hat z)|_{k_z \to 0}.
\end{align}
The spatially dependent part of the stray field only possesses  in-plane components as the reciprocal lattice vectors ${\bf G}_\perp$ are in-plane. 

At a small but finite $k_z$, there exists also a spatially modulated $z$-component of the stray field that is approximately given by $- \frac{M_s}{Q} i k_z  \phi_\alpha({\bf r}_\perp)$ giving rise to a stray field energy density proportional to $k_z (i  \phi_\alpha({\bf r}_\perp) \delta \vec m^*_{\alpha,z}({\bf r}_\perp) + {\rm c.c.})$ that accounts for the integrand of Eq.~\eqref{vinftydip}.
Examples for this integrand are shown in Fig.~\ref{fig:dip}. It is only finite in regions where the spin wave function possesses a finite $z$-component, i.e., the local magnetization oscillates out-of-plane because it derives from a coupling to the $z$-component of the dipolar stray field. The importance of dynamic dipolar interactions for the nonreciprocity of spin excitations has been pointed out before for other non-collinear spin textures\cite{Camley,Otalora,Henry}.

\subsubsection{Numerical values for the velocities $v^{\rm DMI}_{\alpha,\infty}$ and $v^{\rm dip}_{\alpha,\infty}$}

\begin{table}
\begin{tabular}{|c|c|c|c|}
\hline
  & $\frac{v^{\rm DMI}_{\alpha,\infty}}{v_{{\rm FP}, \infty}}$ &  $\frac{v^{\rm dip}_{\alpha,\infty}}{v_{{\rm FP}, \infty}}$ & $\frac{v^{\rm DMI}_{\alpha,\infty}+v^{\rm dip}_{\alpha,\infty}}{v_{{\rm FP}, \infty}}$ \\
\hline
CCW & 0.83 & 0.25 & 1.08 \\
\hline
B & 0.24 & 0.21 & 0.45  \\
\hline
CW & 0.07 & 0.10 & 0.17\\
\hline
\end{tabular}
\caption{\label{table:1} Non-reciprocal velocities of the bulk spin wave dispersion for the CCW, B, and CW mode 
for the magnetic field $H = 0.4 H_{c2}$, that is attributed to the DM interaction, $v^{\rm DMI}_{\alpha,\infty}$, and to the stray field energy, $v^{\rm dip}_{\alpha,\infty}$, see Eqs.~\eqref{vinftyDMI} and \eqref{vinftydip}.}
\end{table}

The numerically evaluated values for the velocities attributed to the DM interaction, $v^{\rm DMI}_{\alpha,\infty}$, and to the stray field energy, $v^{\rm dip}_{\alpha,\infty}$, at the magnetic field $H = 0.4 H_{c2}$ are listed in Table \ref{table:1}. The sum of the two velocities listed in the last column is consistent with the numerically evaluated spectrum shown in Fig.~\ref{fig:SkXspectrum}. Note that this value even exceeds $v_{{\rm FP},\infty}$ for the CCW mode, which is only possible due to the contribution of the stray field energy as $v^{\rm DMI}_{\alpha,\infty} \leq v_{{\rm FP},\infty}$. The dipolar contribution $v^{\rm dip}_\infty$ for the three modes is substantial and corresponds to $23\%$, $47\%$, and $59\%$ for the CCW, breathing and CW mode, respectively.
The strength of $v^{\rm dip}_{\alpha,\infty}$ is weighted by the parameter $\chi_{\rm con}^{\rm int}$, see Eq.~\eqref{vinftydip}, that is larger for Cu$_2$OSeO$_3$, $\chi_{\rm con}^{\rm int} = 1.76$, than for MnSi, $\chi_{\rm con}^{\rm int}= 0.34$ \cite{B20_FMR}. Consequently, the dipolar contribution is particularly important for the material Cu$_2$OSeO$_3$ as $\chi_{\rm con}^{\rm int}$ is relatively large.